\documentclass[aip,jcp,amsmath,amssymb,preprint]{revtex4-1}
\usepackage[utf8]{inputenc}
\usepackage{mfirstuc}
\usepackage{graphicx}
\usepackage{subfigure}
\usepackage{float}
\usepackage{gensymb}
\usepackage{amsmath}
\usepackage{color}
\usepackage{comment}
\usepackage{multirow}
\usepackage{hyperref}

\begin{document}

\title{Is the glassy dynamics same in 2D as in 3D? The Adam Gibbs relation test}

\author{Santu Nath}
\affiliation{Dept. of Physics, IIT Roorkee 47667 Uttarakhand, India.}

\author{Shiladitya Sengupta}
\email{shiladityasg@ph.iitr.ac.in}
\affiliation{Dept. of Physics, IIT Roorkee 47667 Uttarakhand, India.}

\begin{abstract}
It has been recognized of late that even amorphous, glass-forming materials in two dimensions (2D) are significantly affected by Mermin-Wagner type long wavelength thermal fluctuation which is inconsequential in three (3D) and higher dimensions. Thus any study of glassy dynamics in 2D should first remove the effect of such fluctuations. The present work considers the question of whether the role of spatial dimension on glassy dynamics is only limited to such fluctuations, or whether the nature of glassy dynamics is intrinsically different in 2D. We address this issue by studying the relationship between dynamics and thermodynamics within the framework of the Adam-Gibbs (AG) relation and its generalization the Random First Order Transition (RFOT) theory. Using two model glass-forming liquids we find that even after removing the effect of long wavelength fluctuations, the AG relation breaks down in two dimensions. Then we consider the effect of anharmonicity of vibrational entropy - a second factor highlighted recently that can qualitatively change the nature of dynamics. We explicitly compute the configurational entropy both with and without the anharmonic correction. We show that the anharmonic correction reduces the extent of deviation from the AG relation, but even after taking into account its effects, the AG relation still breaks down in 2D. It is also more prominent if one considers diffusion coefficient rather than $\alpha$-relaxation time. Overall, the impact of the anharmonicity of vibration is larger than the long wavelength fluctuation in determining the qualitative relation between timescales and entropy. The extent and nature of deviation from the AG relation crucially depends on the attractive {\it vs.} repulsive nature of the inter-particle interaction. Thus our results suggest that the glassy dynamics in 2D may be intrinsically different from that in 3D. 
\end{abstract}

\maketitle

\section{Introduction}\label{sec:intro}
It is well-known that the nature of phase transition can be qualitatively different in two dimensions (2D) than in three dimensions (3D) \cite{Tarjus2017}. For example, the Mermin-Wagner theorem argues that in 2D, a periodic structure will be unstable against long wavelength thermal deviation from equilibrium positions of atoms \cite{MerminWagner1966, Hohenberg1967, Mermin1968}. It raised doubt whether \emph{crystalline} solids can exist in 2D. Later the KTHNY theory \cite{Kosterlitz2016, Kosterlitz1973, Nelson1979, Young1979, Illing2017} proposed a mechanism for liquid-to-crystal transition on a two dimensional elastic sheet that does not violate the Mermin-Wagner theorem. In this scenario, the translational and the rotational symmetries are broken at different temperatures. To the contrary, in 3D, they are both broken at the same temperature.

Liquids however can also transform into viscous supercooled liquids and eventually into \emph{amorphous} solids, namely glass, by avoiding crystallization. 
Recently Flenner and Szamel \cite{Flenner2015} have found that several hallmark features of glassy dynamics in 3D are different in 2D. The plateaus in the mean squared displacement (MSD) and in the two point density correlation functions are absent for large enough system sizes in 2D; qualitative and quantitative differences in the growth of dynamical heterogeneity and associated correlation length are also observed. 
These observations generated a lot of interest to understand the origin of these apparent differences. Shiba {\it et al.} \cite{Shiba2016} pointed out that in 2D, even \emph{amorphous} matter should exhibit significant long wavelength thermal fluctuations, elucidated in the Mermin-Wagner theorem. It was shown that the effect of these fluctuations can be corrected for by considering the so-called cage-relative coordinates \cite{Russo2015, Shiba2016} in which displacement of a particle is measured with respect to its nearest neighbours instead of from a fixed origin. Both computer simulation \cite{Shiba2016} as well as experimental studies in colloidal systems \cite{Illing2017, Vivek2017} show that after introducing such corrections, dynamical quantities such as the MSD and the time correlation functions behave in qualitatively similar way in both two and three dimensions. 
These works highlight the role of long wavelength fluctuation as a new type of thermal fluctuation unique to 2D \cite{Tarjus2017} and emphasize that this effect should be removed first before studying the glassy dynamics in 2D.

These results provide new insights about the dimension dependence of dynamics in glass-forming liquids. However they also raise fundamental questions that beg clarification. Since these previous studies focused on the \emph{empirical description} of glassy dynamics, one wonders whether the role of spatial dimension is fully explained by the long wavelength fluctuations, or whether the two dimensional glass-formers  are intrinsically different from the three dimensional ones. This question is particularly significant for  entropy-based glass transition theories, namely the Adam Gibbs (AG) relation \cite{Gibbs1958, Adam1965, Debenedetti2021} and its generalization, the Random First Order Transition (RFOT) theory \cite{Kirkpatrick2015, Lubchenko2007, Bouchaud2004, Biroli2012, Starr2013, Karmakar2015}. In these theories one first writes an expression for the free energy barrier for structural relaxation which depends on entropy, and thence estimates the structural relaxation time. The AG relation between the structural relaxation time $\tau_\alpha$ and the configurational entropy $S_c$ at a given temperature $T$ is given by,
\begin{align}
    \tau_\alpha = \tau_\infty \, \exp(\frac{A}{TS_c}),
    \label{eqn:AG}
\end{align}
where $A$ is an activation free energy barrier and $\tau_\infty$ is the limitting high temperature value of relaxation time. Eqn. \ref{eqn:AG} was originally introduced for polymer glass models in 3D, and the spatial dimension dependence is not explicit. To the contrary, the RFOT theory introduces a characteristic spatial correlation lengthscale $\xi$, so an explicit dimension dependence is plausible. However, in the RFOT framework, the lengthscale $\xi$ is related to timescale and configuration entropy by a set of exponents which are {\it a priori unknown}. So an explicit dimension dependence is not guaranteed. In particular, whether the AG relation (Eqn. \ref{eqn:AG}) is recovered as a special case of the RFOT relations, depends on the exponent values. Previously some of us found that the AG relation breaks down in 2D, while it remains valid in three and higher spatial dimensions \cite{Sengupta2012}. Given the recent discovery of the importance of Mermin-Wagner type excitations in 2D, the question naturally arises whether the observed deviation is a consequence of such low frequency vibrations, or whether spatial dimension has an intrinsic effect on the glass transition. 

The long wavelength fluctuation affects dynamics (relaxation time), but not thermodynamics (configurational entropy), by definition. Within the potential energy landscape (PEL) formalism, the configurational entropy is usually computed by neglecting the anharmonic contribution to vibrational entropy \cite{Sciortino2005, Nandi2022, Das2022}, see Sec. \ref{sec:Defn} for details. However, recently it has been observed that the anharmonic contribution is significant in the type of glass-formers we study here, and may even change the qualitative behaviour of a system \cite{Das2022}. 

Thus the goal of the present study is to critically examine the validity of the AG relation in 2D, after explicitly taking into account the effects of (i) Mermin-Wagner type fluctuations that affect dynamics and (ii) the anharmonic nature of vibration entropy that affect thermodynamics. 
The rest of the paper is divided into the following sections: Sec. \ref{sec:Defn} defines the relevant quantities. Model glass-formers and simulation details are described in Sec. \ref{sec:simuDet}. Sec. \ref{sec:results} discusses the results of our analysis. Finally in Sec.\ref{sec:conclude} we summarize the results and conclude.

\begin{figure*}[htp!]
    \centering
    \includegraphics[keepaspectratio, width=0.32\textwidth]{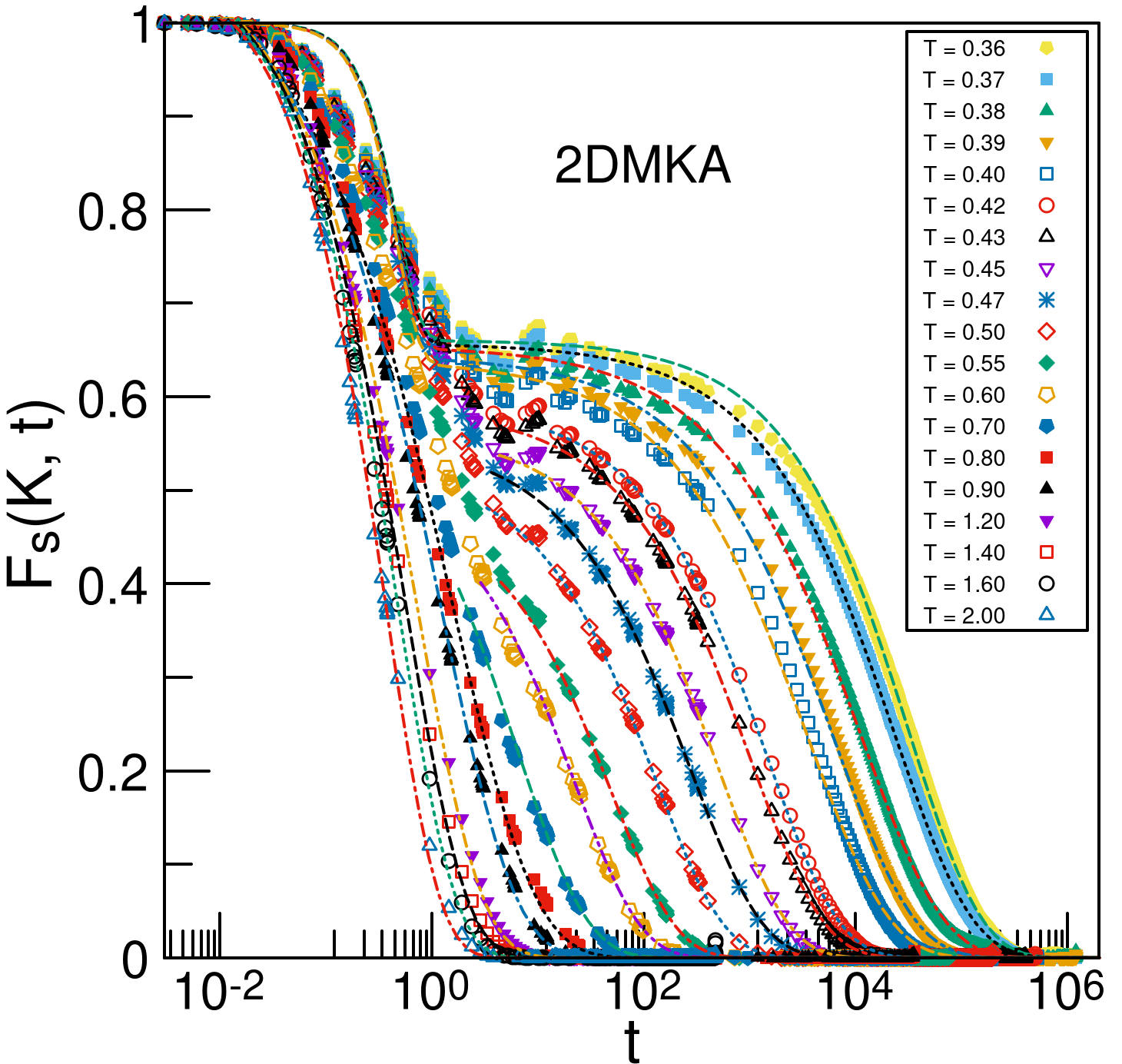}
    \includegraphics[keepaspectratio, width=0.32\textwidth]{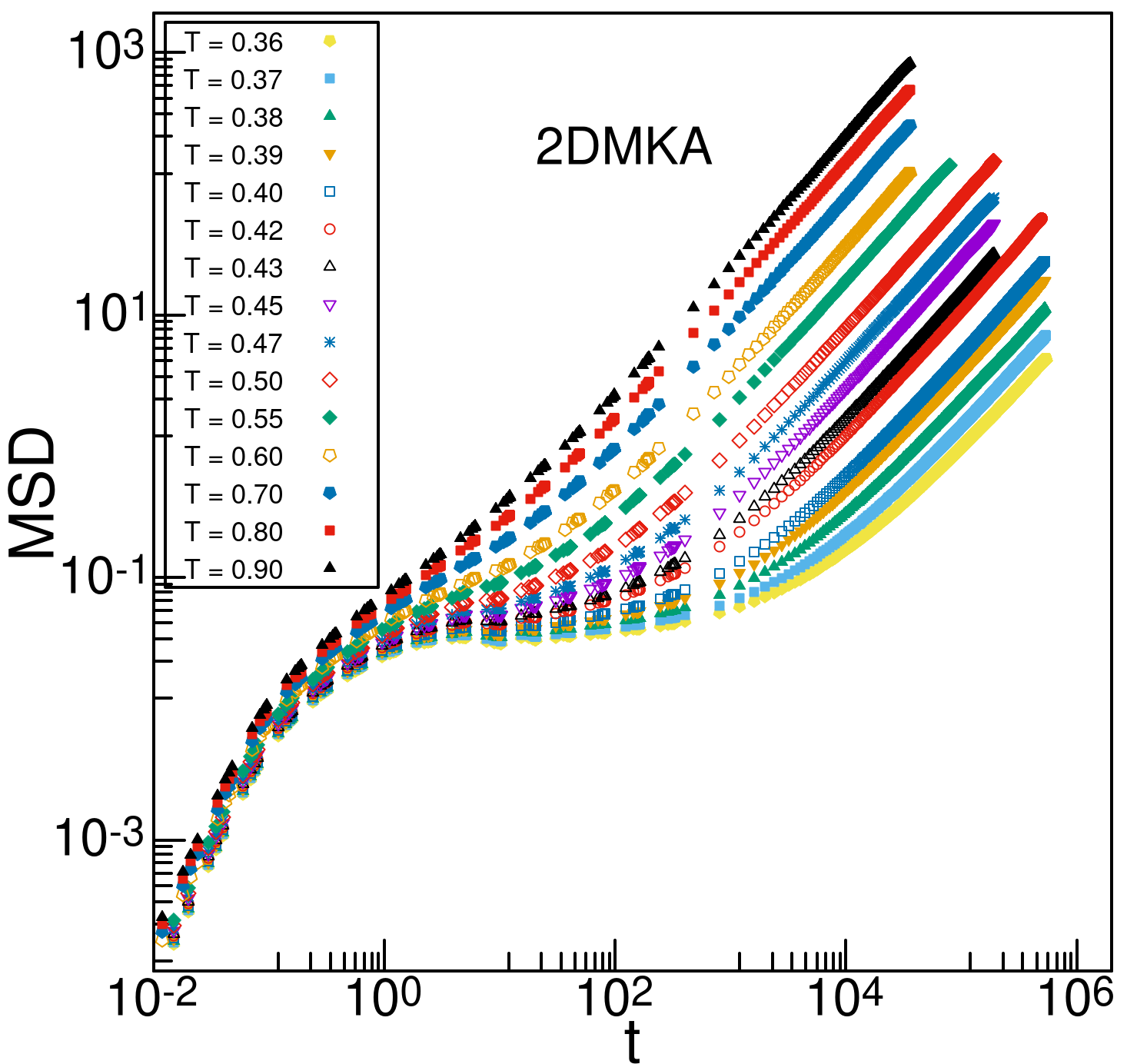}\\
    \includegraphics[keepaspectratio, width=0.32\textwidth]{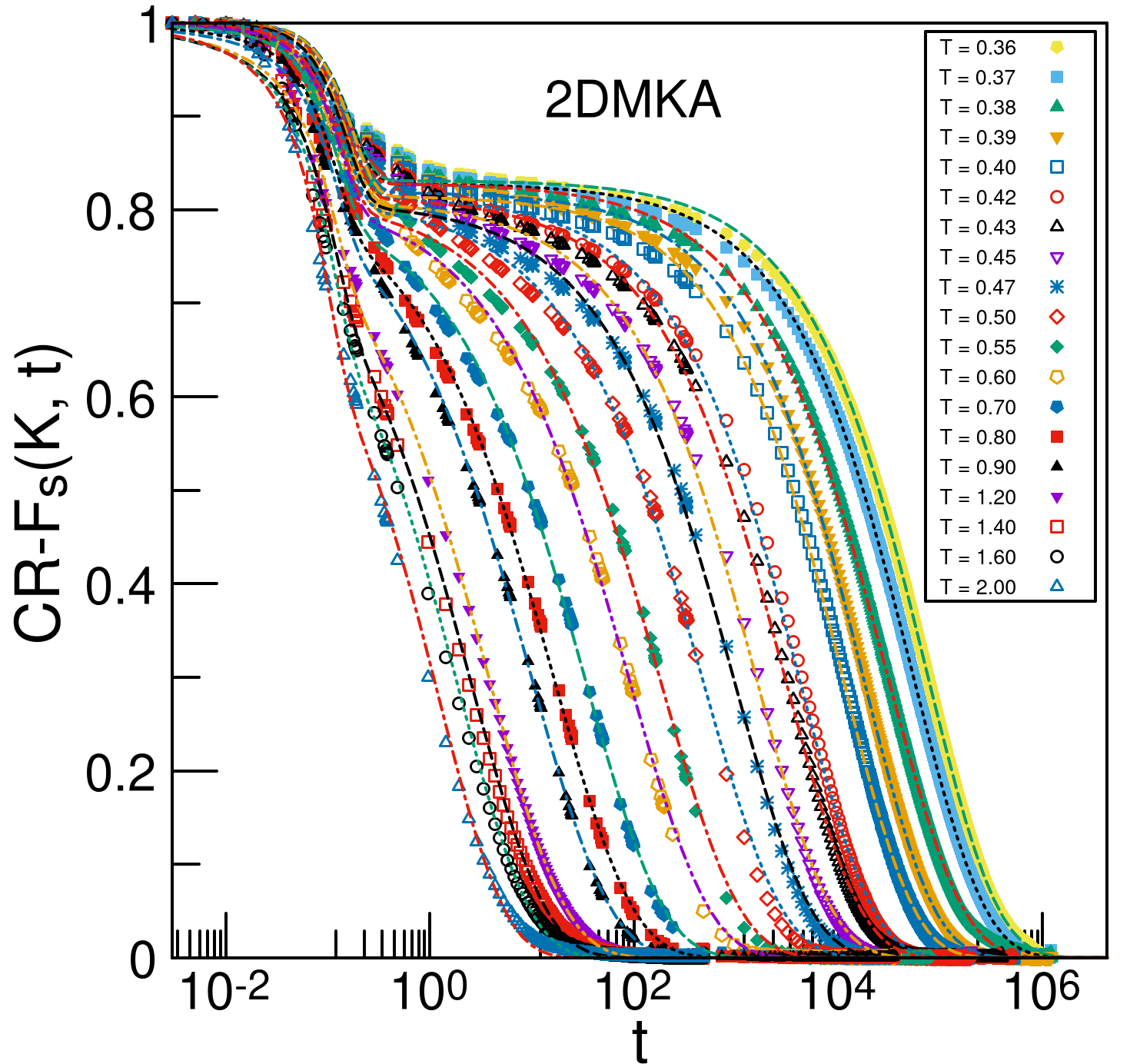}
    \includegraphics[keepaspectratio, width=0.32\textwidth]{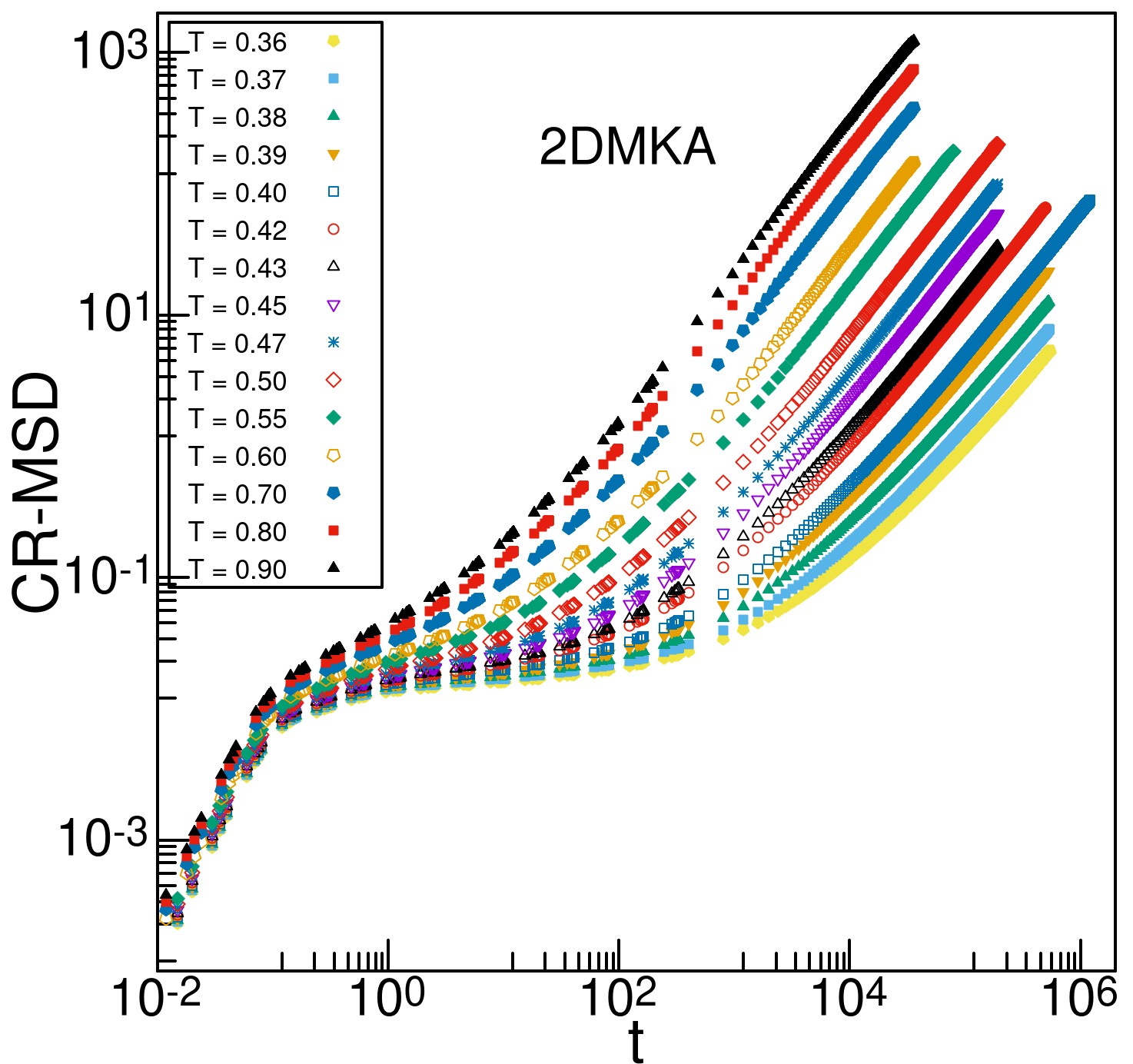}
    \caption{Temperature dependence of the $F_s(K, t)$, MSD (top row) and CR-$F_s(K, t)$, CR-MSD (bottom row) for the 2DMKA model. Lines are fits to Eqn. \ref{eqn:tau_alpha}.}
    \label{fig:Fskt2DMKA}
\end{figure*}

\begin{figure*}[htp!]
    \centering
    \includegraphics[keepaspectratio, width=0.32\textwidth]{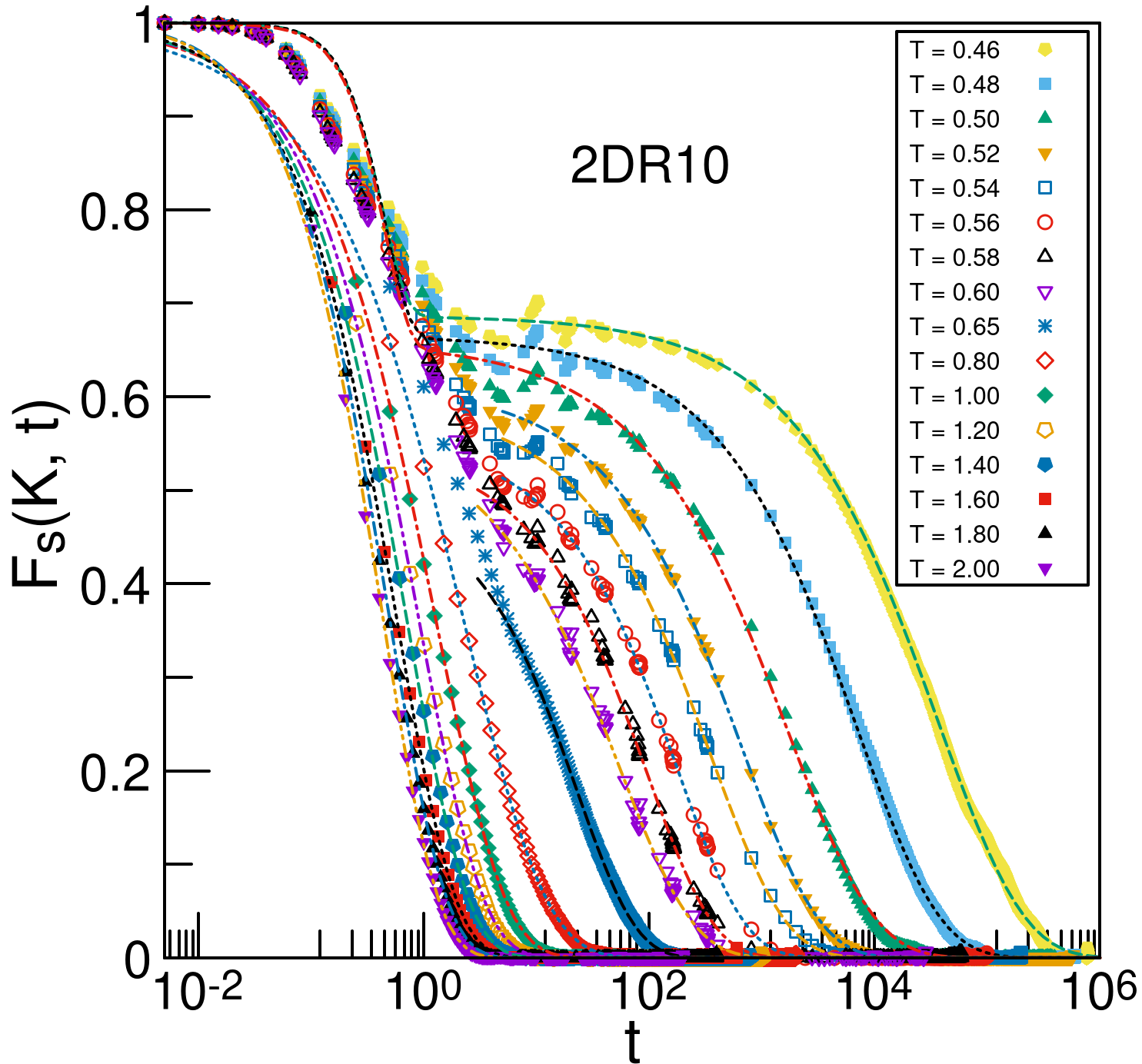}
    \includegraphics[keepaspectratio, width=0.32\textwidth]{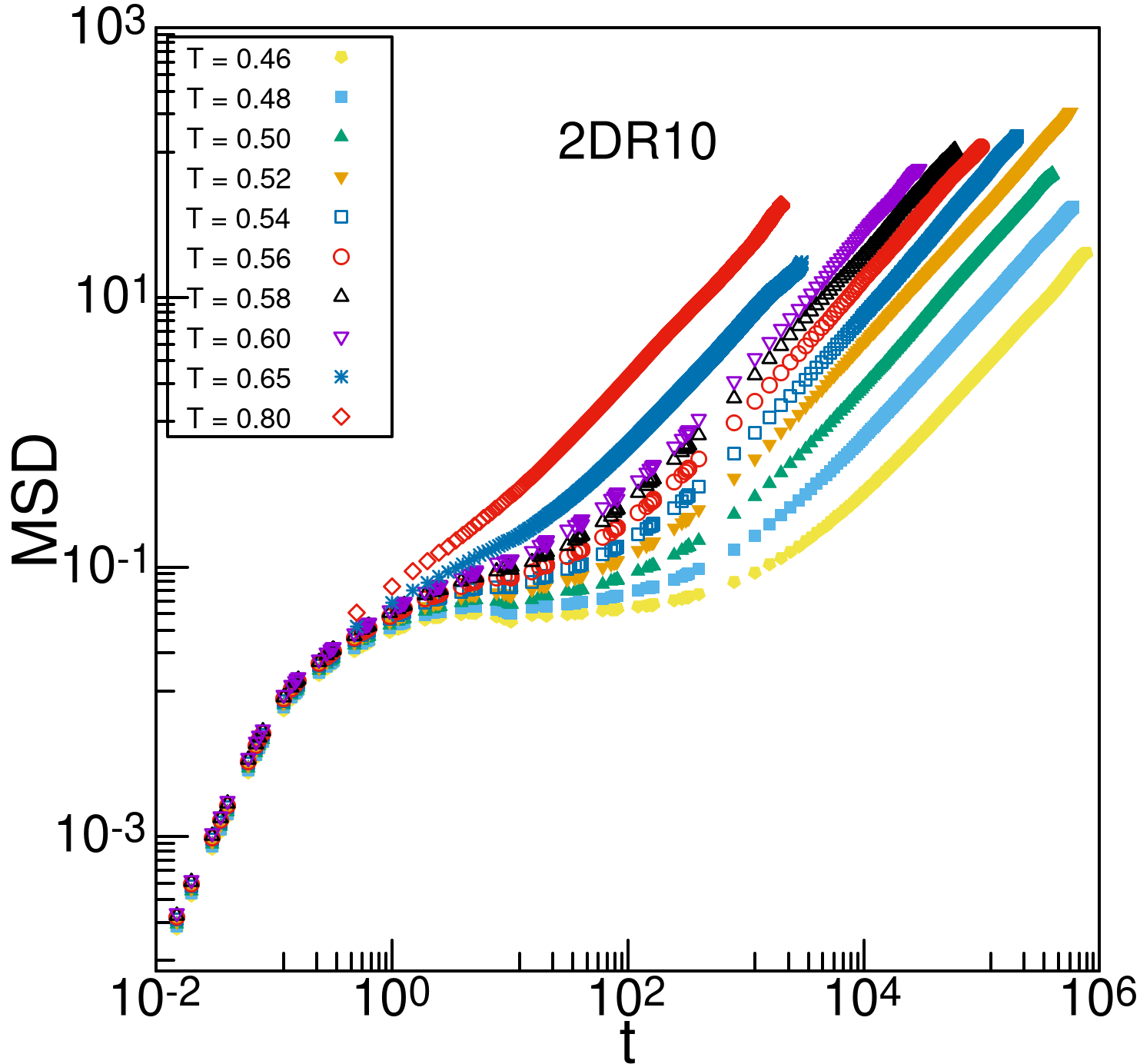}\\
    \includegraphics[keepaspectratio, width=0.32\textwidth]{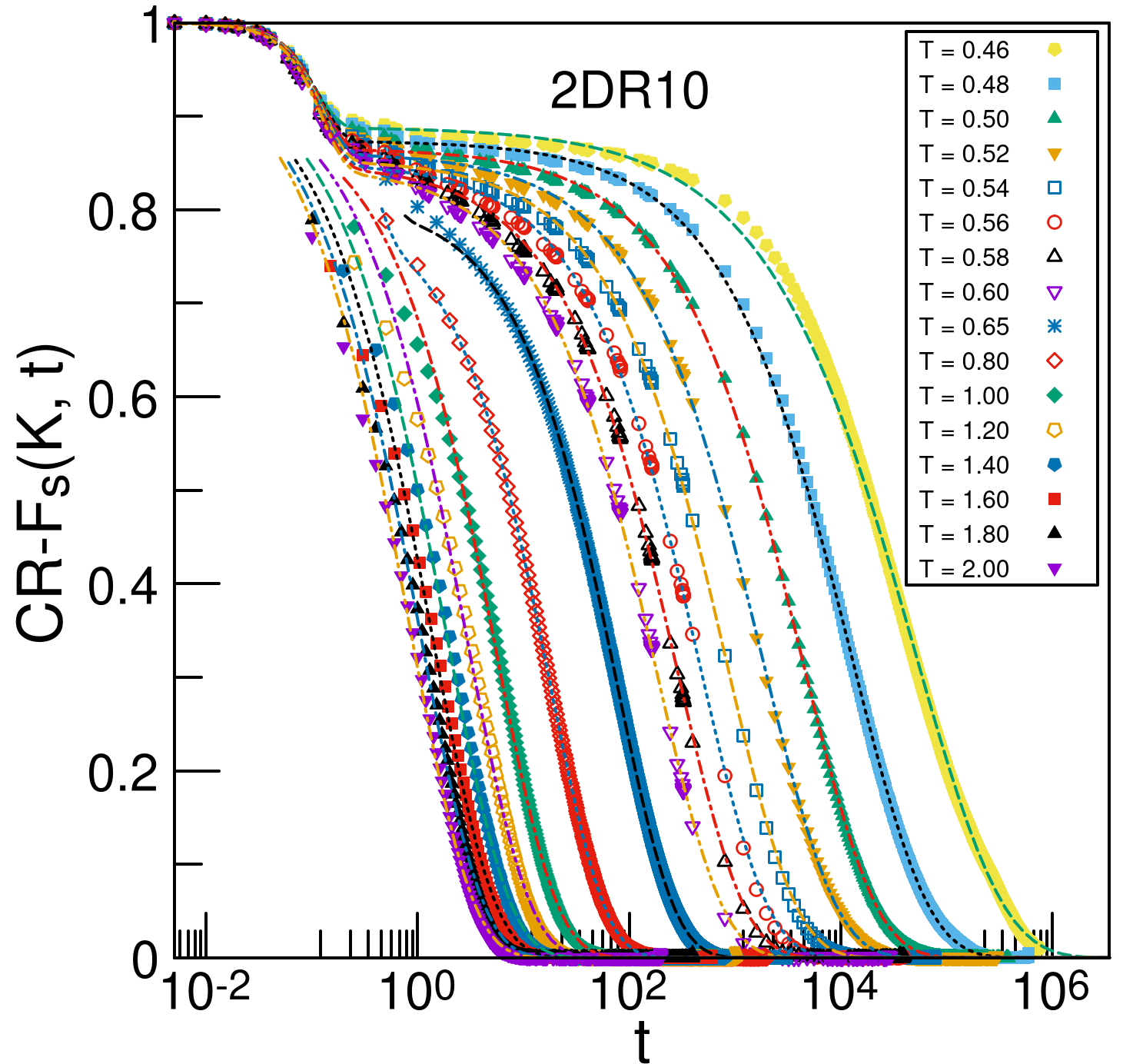}
    \includegraphics[keepaspectratio, width=0.32\textwidth]{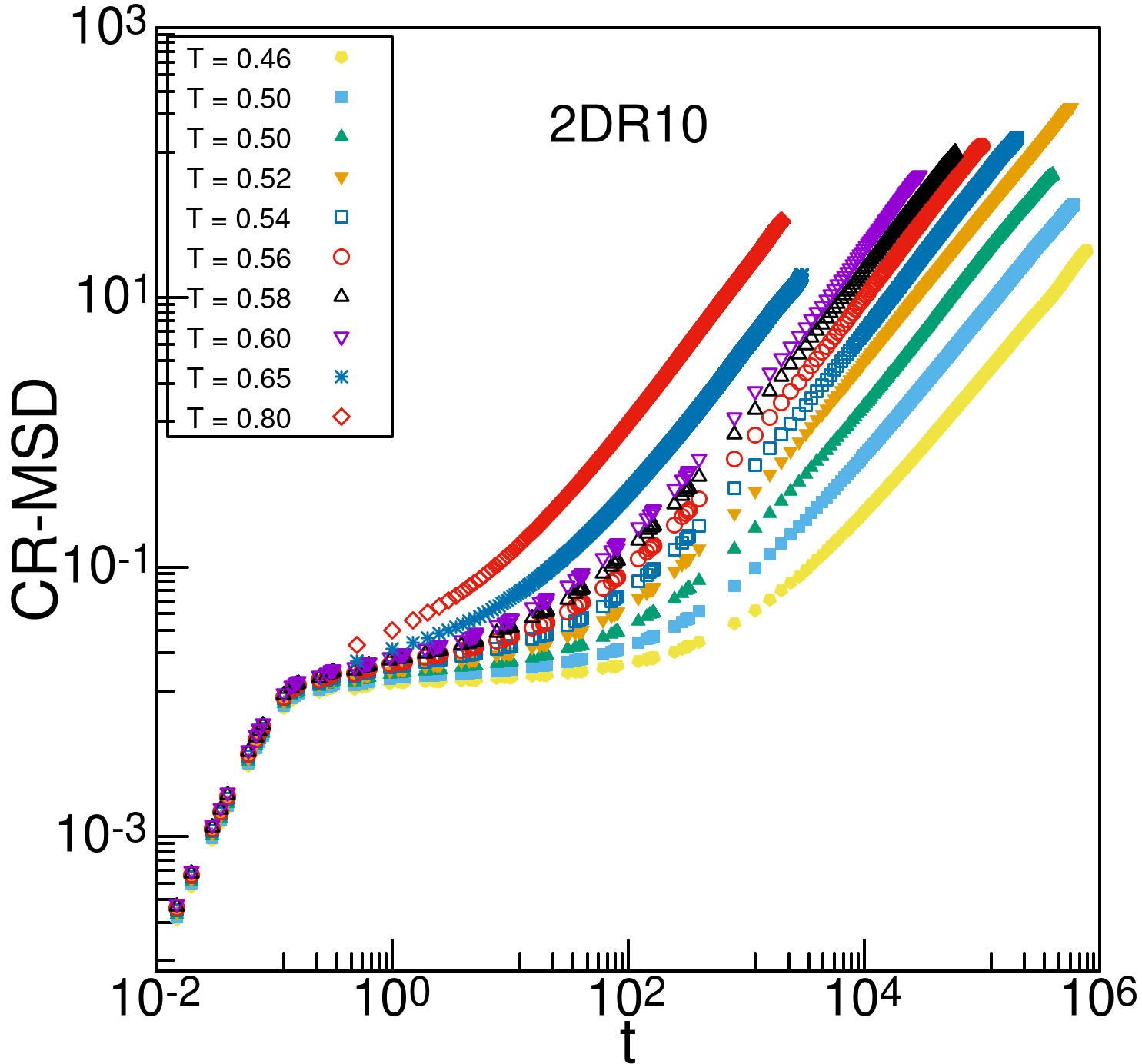}
    \caption{Temperature dependence of the $F_s(K, t)$, MSD (top row) and the CR-$F_s(K, t)$, CR-MSD (bottom row) for the 2DR10 model. Lines are fits to Eqn. \ref{eqn:tau_alpha}.}
    \label{fig:Fskt2DR10}
\end{figure*}

\section{Definitions}\label{sec:Defn}
\subsection{Dynamics}

\paragraph{\textbf{Cage relative (CR) displacement}} To correct for the long wavelength fluctuation in two dimensions, we compute the cage relative (CR) displacement of a particle $i$ in a time interval $t$ \cite{Shiba2016, Russo2015}:
\begin{equation}
    \Delta \vec{r}_{i, CR}(t) = \Delta \vec{r}_{i}(t) - \frac{1}{N_{nn}(i)} \sum_j  \Delta \vec{r}_{j}(t),
    \label{eq:CRD}
\end{equation}
where $N_{nn}(i)$ is the number of nearest neighbours of $i$, at $t=0$, forming a transient cage around it. The right hand side of Eqn. \ref{eq:CRD}, contains two terms. The first term $\Delta \vec{r}_i (t) = \vec{r}_i (t+t_0) - \vec{r}_i (t_0)$ describes the actual displacement of the $i$-th particle in time interval $t$, with respect to a time origin $t_0$. The second term represents the displacement of the center of mass of the nearest neighbours of particle $i$ at time $t=0$. Thus Eqn. \ref{eq:CRD} measures the displacement of the particle $i$ relative to the center of mass of its nearest neighbours at $t=0$.

\paragraph{\textbf{Time correlation functions}}
We analyze the dynamics using two point time correlation functions of instantaneous local density $\rho(\vec{r},t) = \sum_{i=1}^N \delta \left(\vec{r} - \vec{r_i}(t)\right)$ of $N$ point particles. The standard self intermediate scattering function $F_s(\vec{K}, \tau)$ is defined as,
\begin{align}
    F_s(\vec{K}, \tau) &= \frac{1}{N} <\rho(\vec{K},t_0+\tau)\rho(-\vec{K},t_0)> \nonumber\\
    &= \frac{1}{N}  \sum_{i=1}^N < e^{-i\vec{K}.[\vec{r_i}(t_0+\tau) - \vec{r_i}(t_0)]} > \nonumber\\
    &= \frac{1}{N}  \sum_{i=1}^N < e^{-i\vec{K}.\vec{r_i}(t_0+\tau)} \times e^{i\vec{K}.\vec{r_i}(t_0)} >
    \label{eq:Fskt}  
\end{align}
where $\rho (\vec{K},t) =  \int d\vec{r}\, \rho(\vec{r},t)  e^{-i\vec{K}.\vec{r}}  = \sum_{i=1}^N e^{-i\vec{K}.\vec{r_i}(t)}$ is the Fourier transform of $\rho(\vec{r},t)$. $\langle \cdot \rangle \equiv \frac{1}{N_{t_0}} \sum_{t_0=1}^{N_{t_0}} \cdot $ represents averaging over different time origins $t_0$. Since the liquid is isotropic, the correlation function depends only on the magnitude of the wave vector $\vec{K}$. Thus we further do a averaging over different directions, to compute a correlation function $F_s(K, \tau)$ that depends only on the magnitude $K = |\vec{K}|$. We chose $K$ at the first peak of the static structure factor $S(K)$.  

In addition, we report data obtained from another time correlation function - the overlap function. The details are described elsewhere \cite{Sengupta2012}.

We define cage-relative self-intermediate scattering function (CR-$F_{s}(K,t)$) as the time correlation function using the cage relative displacement replacing standard displacement in Eqn. \ref{eq:Fskt}. 

\vspace{3mm}
\paragraph{\textbf{$\alpha$-relaxation time ($\tau_{\alpha}$)}}
From time correlation functions $F_s(K,t)$ and CR-$F_s(K,t)$, the structural relaxation time $\tau_{\alpha}$ characterizing the dynamics of standard and cage-relative displacement fields is estimated by fitting $y(t)$ to the following functional form:
\begin{equation}
    y(t) = (1 - f_c)\, e^{-(t/\tau_s)^{2}} + f_c \, e^{-(t/\tau_\alpha)^{\beta}}
    \label{eqn:tau_alpha}
\end{equation}
where $y(t)$ denotes the appropriate time correlation function. The first term describes the short time decay and the second term the long time $\alpha$ decay of the correlation function. $f_c, \tau_s$ and $\beta$ denote the plateau height, the characteristic time scale for short time decay and the Kohlrausch-Williams-Watts exponent respectively. In addition, we also estimate $\tau_{\alpha}$ from the overlap function of standard displacement field the details of which is described in Ref. \cite{Sengupta2012}.

\vspace{3mm}
\paragraph{\textbf{Mean Squared displacement}}
Standard mean squared displacement (MSD) and its cage-relative generalization CR-MSD are defined as
\begin{align}
    \mbox{MSD} (t) &= \langle \frac{1}{N} \sum_{i=1}^{N}[\Delta \vec{r}_i (t)]^2 \rangle\nonumber\\
    \mbox{CR-MSD} (t) &= \langle \frac{1}{N} \sum_{i=1}^{N}[\Delta \vec{r}_{i,CR} (t)]^2 \rangle
    \label{eq:MSD}
\end{align}
where $\Delta \vec{r}_i (t), \Delta \vec{r}_{i,CR} (t)$ are defined in Eqn. \ref{eq:CRD}. $\langle \cdot \rangle$ denotes the ensemble average. 

\vspace{3mm}
\paragraph{\textbf{Diffusion coefficient}}
We estimate the diffusion coefficient from the MSD and CR-MSD using the definition
\begin{align}
    D &= \lim_{t \rightarrow \infty} \frac{\mbox{MSD} (t)}{4\,t} \nonumber\\
    D_{CR} &= \lim_{t \rightarrow \infty} \frac{\mbox{CR-MSD} (t)}{4\,t}
    \label{eq:D}
\end{align}

\subsection{Thermodynamics}

\subsubsection{Configurational entropy $S_c$ in harmonic approximation}
Within the potential energy landscape formalism \cite{Sciortino2005}, the configurational entropy $S_c (T)$ is a measure of the number of independent potential energy minima that a system can sample at a given temperature $T$ and a given density $\rho$. It is defined as the ensemble-averaged difference between the total and the vibrational entropy:
\begin{align}
    S_c &\equiv S_{total}(T) - S_{vib}(T) \nonumber\\
    S_c^{har} (T) &= S_{total}(T) - S_{vib}^{har}(T).
    \label{eqn:schar}
\end{align}
where $S_{vib}^{har}$ indicates that the vibrational entropy is computed using the the harmonic approximation \cite{Sengupta2011, Nandi2022}. 

The total entropy of the liquid $S_{total}$ at a given temperature and density is computed using thermodynamic integration method, the details of which is described elsewhere \cite{Sengupta2011,Nandi2022}.

\subsubsection{(Vibrational) potential energy (PE)}
\paragraph{\textbf{Harmonic approximation}} At a given temperature $T$, the average per-particle (vibrational) potential energy in the harmonic approximation \textcolor{black}{in 2D} can be written as (setting the Boltzmann constant $k_B=1$),
\begin{equation}
    U_{har}(T) = \langle U_{IS}(T) \rangle + \textcolor{black}{\frac{2}{2}}T
    \label{eq:Uhar}
\end{equation}
Here $\langle U_{IS} \rangle$ is the per particle potential energy at a minimum (inherent stricture, IS). We explicitly show the $\langle \cdot \rangle$ to emphasize that it is ensemble averaged.

\paragraph{\textbf{Anharmonic correction}} The average anharmonic contribution $U_{anhar}(T)$ to the (vibrational) per-particle potential energy can be found by subtracting the harmonic component of potential energy from the full potential energy \cite{Sciortino2005, Nandi2022, Das2022}, {\it i.e.} 
\begin{align}
    U_{anhar} &\equiv U - U_{har}
    \label{eq:Uanhar}
\end{align}

We have computed $U_{anhar} (T)$, using two methods differing in the averaging procedure, leading to two slightly different estimates of $S_c$. The methods are described below. 

\subsubsection{Anharmonic vibrational entropy: method 1}
\paragraph{\textbf{Anharmonic correction to (vibrational) PE}}
$U_{anhar}(T)$ is computed using the following procedure:
\begin{align}
    U_{anhar}(T)&= \langle U(T) \rangle - \langle U_{IS}(T) \rangle - \textcolor{black}{\frac{2}{2}}T \nonumber\\
                &= \sum_{j=2}^{j_{max}} C_j T^j
    \label{eq:uanh1}
\end{align}
where we used Eqn. \ref{eq:Uhar}. Here $\langle U(T) \rangle$ is the average potential energy per particle at a temperature $T$. The second line describes the temperature dependence of $U_{anhar}$ by a fit polynomial with $C_j$'s being the unknown fit parameters \cite{Sciortino2005}.

\vspace{3mm}
\paragraph{\textbf{Anharmonic vibrational entropy}} From Eqn. \ref{eq:uanh1}, the anharmonic component of vibrational entropy can be written as,
\begin{align}
    S_{vib}^{anhar}(T) &= \int_0^T dT' \frac{1}{T'} \frac{\partial U_{anhar}(T')}{\partial T'} = \sum_{j = 2}^{j_{max}} \frac{j}{j - 1} C_j \, T^{j - 1}
    \label{eq:sanh1}
\end{align}

\subsubsection{Anharmonic vibrational entropy: method 2}
\paragraph{\textbf{Anharmonic correction to (vibrational) PE}}
In this method the (vibrational) potential energy in a given basin at a given temperature $T$ is directly estimated. Starting from inherent structures obtained by minimizing configurations equilibrated at a parent temperature $T_p$, short molecular dynamics (MD) runs of duration $\approx$ MSD plateau time are performed at different target temperatures $T \leq T_p$. The anharmonic contribution is then computed from 
\begin{align}
    U_{anhar}(T,T_p) &= \langle U(T) - U_{IS}(T_p) \rangle - \textcolor{black}{\frac{2}{2}}T \nonumber\\
    &= \sum_{j=2}^{j_{max}} C_j (T_p)\, T^j
    \label{eq:uanh2}
\end{align}
Here $C_j (T_p)$'s are $T_p$ dependent unknown fit parameters.
\vspace{3mm}
\paragraph{\textbf{Anharmonic vibrational entropy}}  In method 2, the anharmonic contribution to vibrational entropy is computed from Eqn. \ref{eq:uanh2} as, 
\begin{align}
    S_{vib}^{anhar}(T_p) &= \int_0^{T_{p}} dT' \frac{1}{T'} \frac{\partial U_{anhar}(T',T_p)}{\partial T'} \nonumber\\
    &= \sum_{j = 2}^{j_{max}} \frac{j}{j - 1} C_j (T_p)\, T_p^{j - 1}
    \label{eq:sanh2}
\end{align}

\subsubsection{Anharmonic configurational entropy} Finally, the configurational entropy with the anharmonic correction to vibrational entropy is given by,
\begin{equation}
    S_c^{anhar} (T) = S_{total}(T) - \bigg(S_{vib}^{har}(T) + S_{vib}^{anhar}(T)\bigg)
    \label{eqn:scanhar}
\end{equation}

\begin{figure*}[htp!]
    \centering
    \includegraphics[keepaspectratio, width=0.28\textwidth]{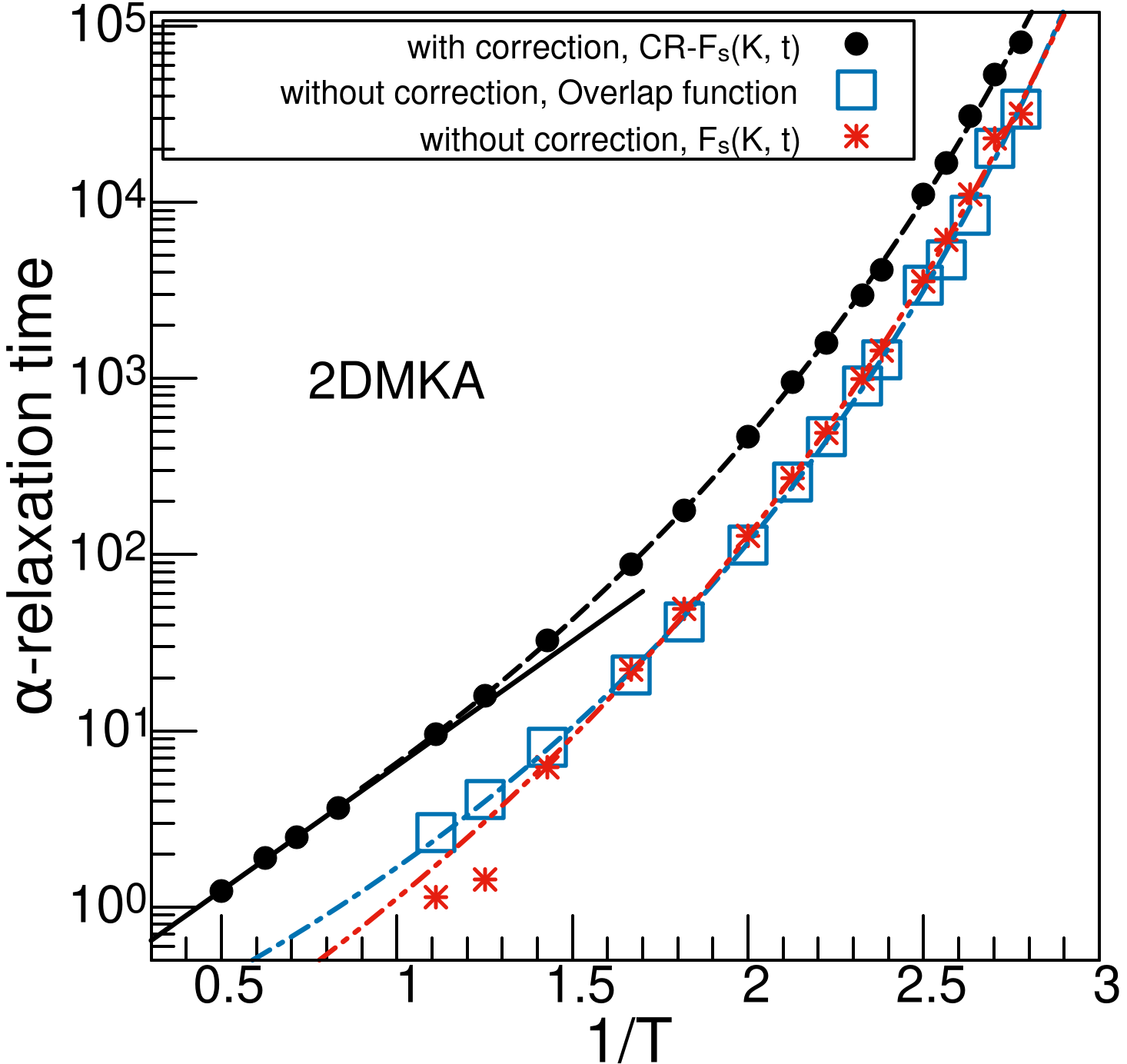}
    \includegraphics[keepaspectratio, width=0.28\textwidth]{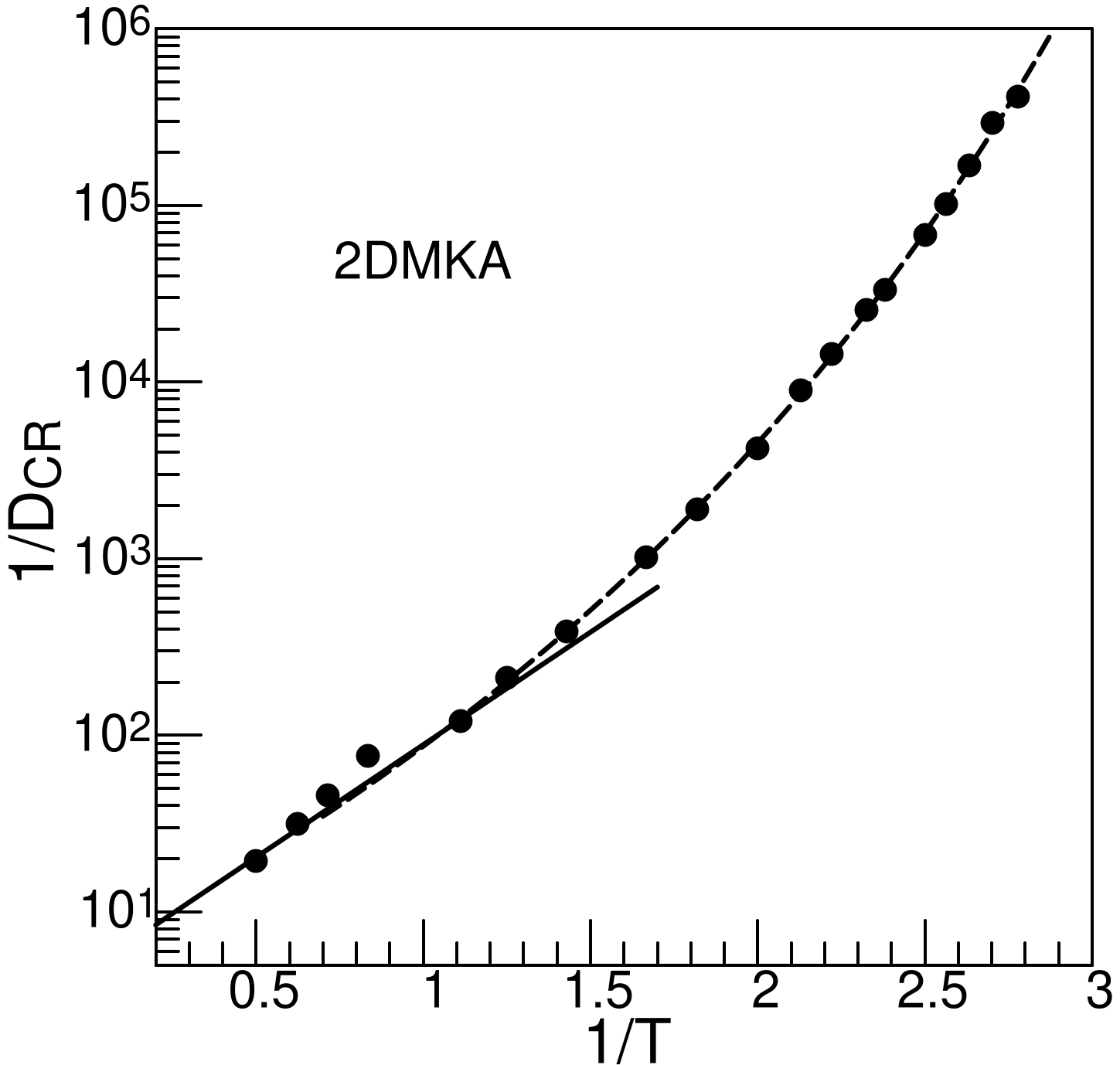}
    \includegraphics[keepaspectratio, width=0.28\textwidth]{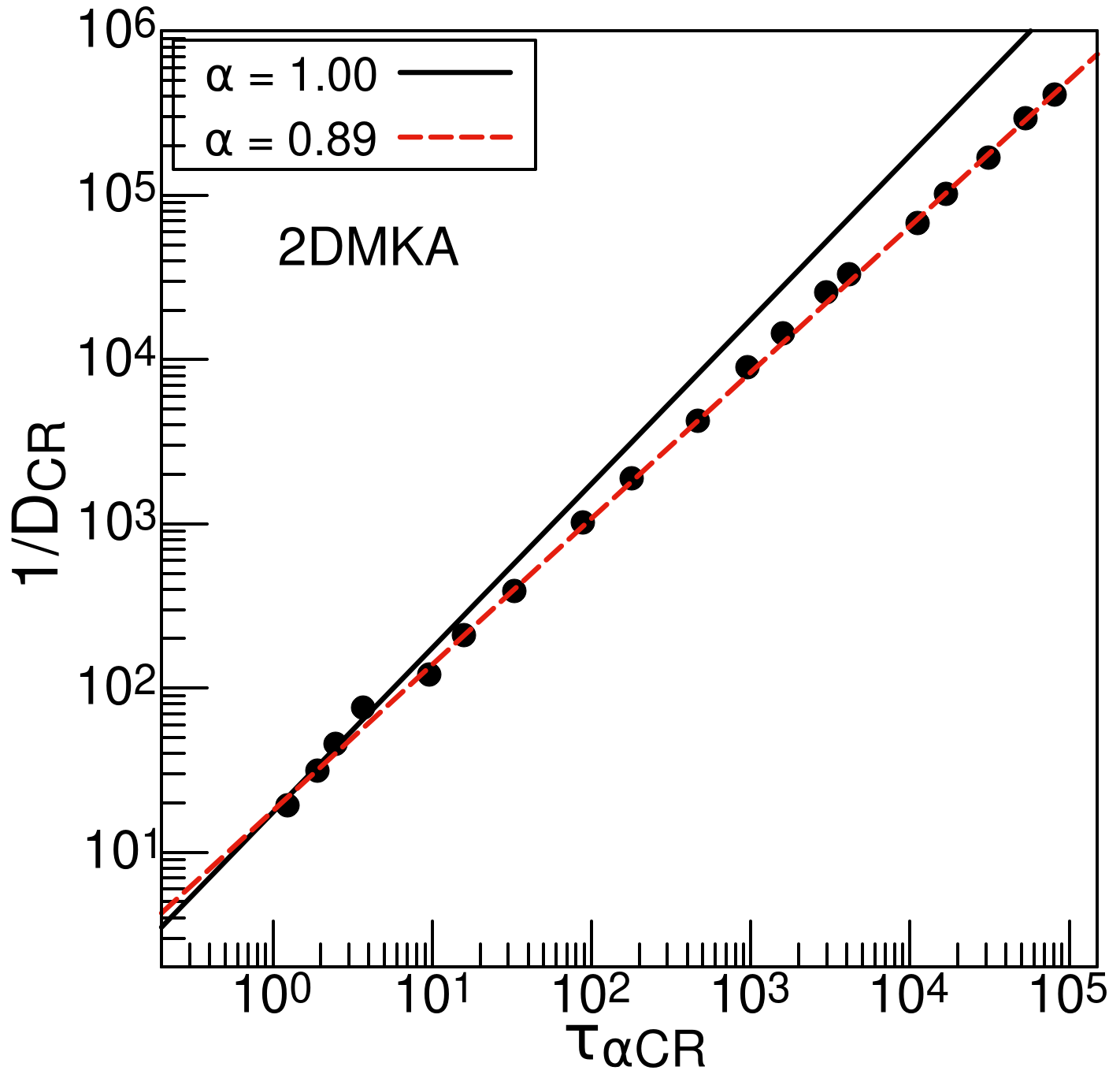}\\
    \includegraphics[keepaspectratio, width=0.28\textwidth]{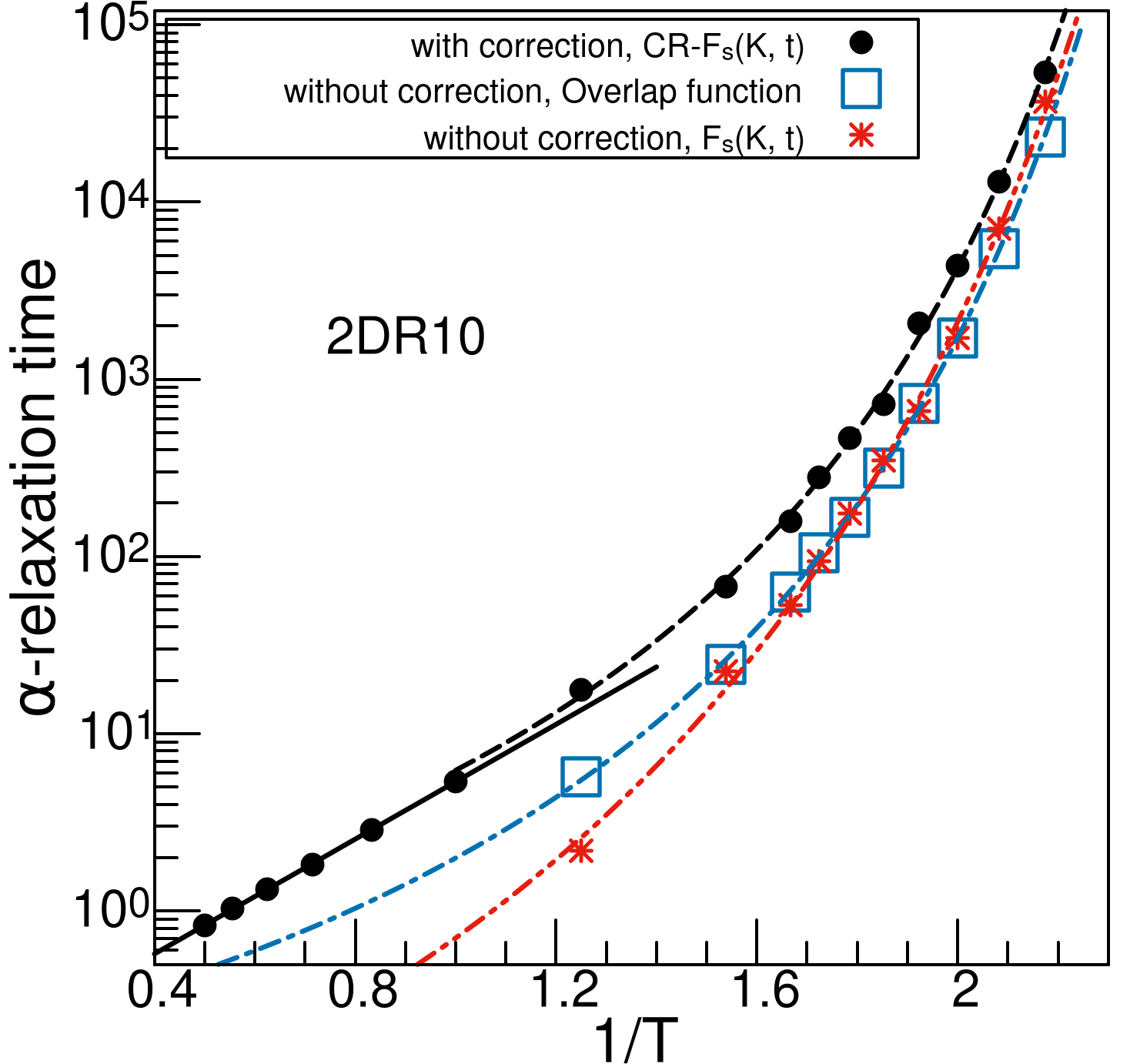}
    \includegraphics[keepaspectratio, width=0.28\textwidth]{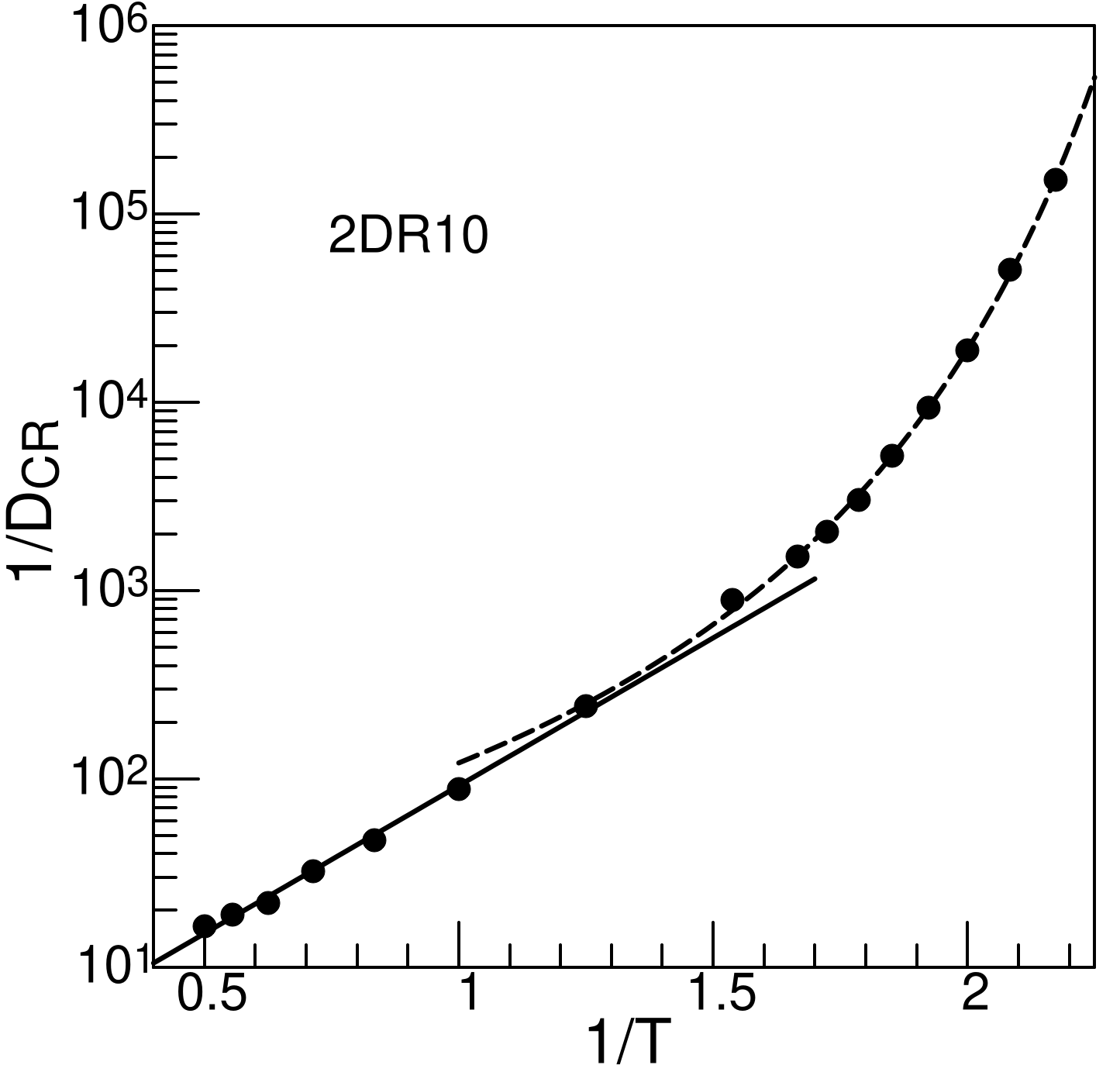}
    \includegraphics[keepaspectratio, width=0.28\textwidth]{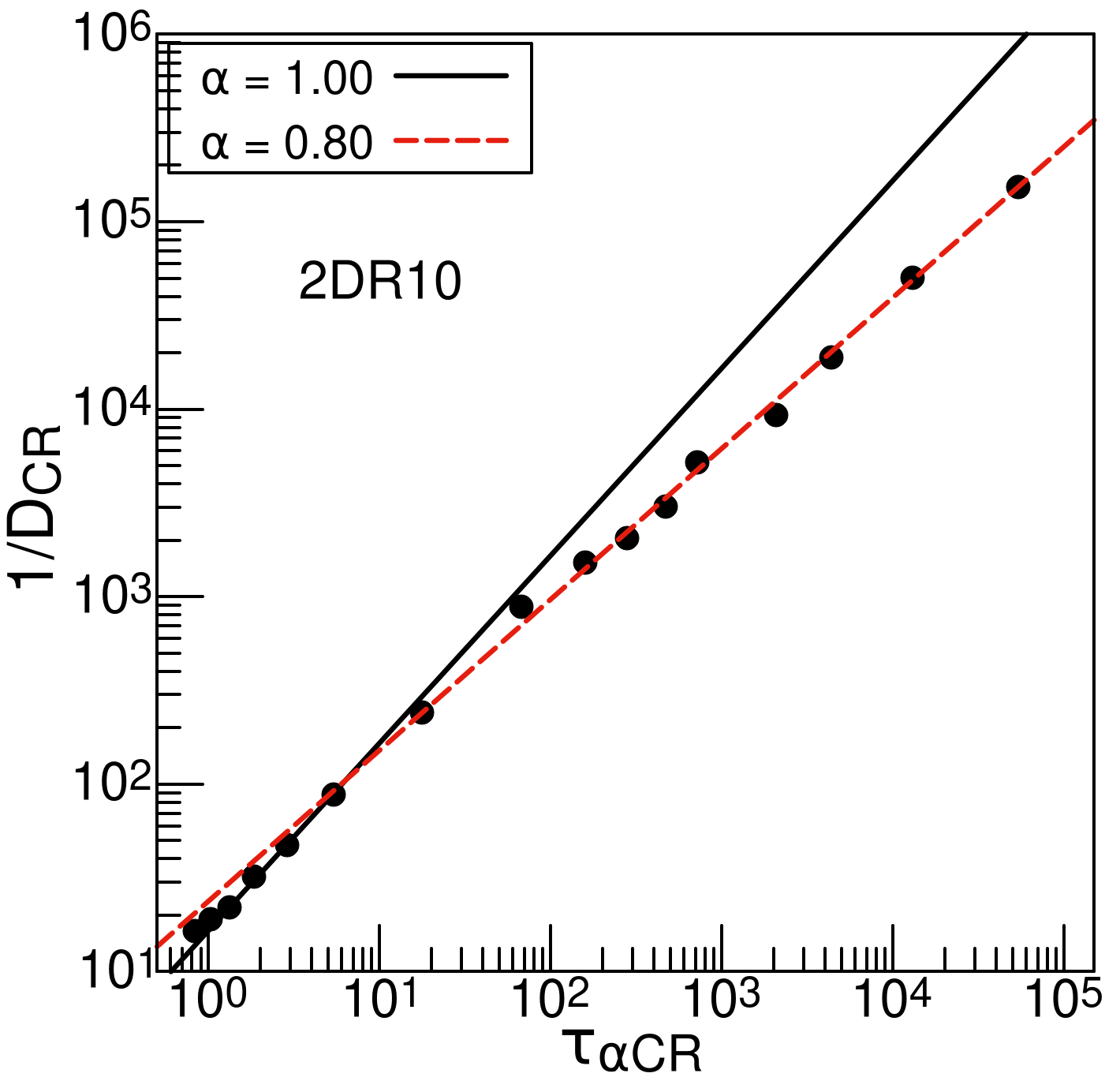}
    \caption{\emph{Effect of long wavelength fluctuations on characteristic timescales.} Inverse temperature dependence of characteristic timescales are shown for the 2DMKA (top row) and the 2DR10 (bottom row) respectively. Time scales are (a) $\tau_{\alpha}$ computed from the standard intermediate scattering function as well as from the standard overlap function (reported in Ref. \cite{Sengupta2012}), (b) $\tau_{\alpha CR}$ estimated from the cage-relative intermediate scattering function, and (c) diffusion coefficient $D_{CR}$ obtained from the cage-relative MSD. Data at low temperatures are fitted to the VFT equation: $\tau_{\alpha} = \tau_0 \exp\left(\frac{D}{T-T_0}\right)$, $D_{CR}^{-1} = (D_{CR}^{-1})_0 \exp\left(\frac{D}{T-T_0}\right)$ while the high $T$ data are fitted to the Arrhenius equation $\tau_{\alpha} (T) = \tau_\infty \exp(\frac{E_0}{k_B\,T})$. Fit parameters are listed in Table \ref{table:taufitParam}. \textcolor{black}{$\tau_{\alpha}$ from the overlap function are scaled by a constant factor of 0.475 (2DMKA) and 0.043 (2DR10) to match with $\tau_{\alpha}$ values from $F_s(K, t)$.} \emph{Stokes-Einstein Breakdown:} On the right most column, comparison of the diffusion time scale $D_{CR}$ and the $\alpha$-relaxation time scale $\tau_{\alpha, CR}$ from the cage-relative displacement show that they are coupled at high temperature, but gets decoupled at low temperatures, indicating breakdown of the Stokes Einstein relation. Taken together these data show that long wavelength fluctuation makes the $\alpha$-relaxation time longer, but the qualitative behaviour of the dynamics is not affected.}
    \label{fig:compareTime}
\end{figure*}

\begin{table*}[htp]
    \centering
    \caption{Parameters for fit to the high temperature Arrhenius relation: $\tau_\alpha (T) = \tau_\infty \, \exp(E_0 / k_B T)$ and the low temperature VFT formula: $\tau_\alpha (T) = \tau_0 \, \exp({D \over T - T_0})$ for 2DMKA and 2DR10 models.}
    \begin{tabular}{|c|c|c|c|c|c|c|c|c|}
    \hline
    Model & Correlation function & Fit function & Temperature range & $\tau_\infty$ & $E_0$ & $\tau_0$ & D & $T_0$ \\
    \hline
    \multirow{4}{*}{2DMKA}  &  CR-$F_s(K,t)$ &  Arrhenius & 0.80 - 2.00 & 0.24 & 3.26 & - & - & - \\
    \cline{2-9}
            & CR-$F_s(K,t)$  & VFT & 0.36 - 0.80 & - & - & 0.42 & 2.27 & 0.18 \\
    \cline{2-9}
            & overlap function & VFT & 0.36 - 0.80 & - & - & 0.12 & 2.11 & 0.19 \\
    \cline{2-9}
            & $F_s(K,t)$  & VFT & 0.36 - 0.80 & - & - & 0.05 & 2.66 & 0.16 \\
    \hline
    \multirow{4}{*}{2DR10}  &   CR-$F_s(K,t)$   &  Arrhenius & 0.80 - 2.00 & 0.13 & 3.73 & - & - & - \\
    \cline{2-9}
       &   CR-$F_s(K,t)$  & VFT & 0.46 - 0.80 & - & - & 0.62 & 1.56 & 0.32 \\
    \cline{2-9}
       & overlap function & VFT & 0.46 - 0.80 & - & - & 0.17 & 1.65 & 0.32 \\
    \cline{2-9}
       &   $F_s(K,t)$  & VFT & 0.46 - 0.80 & - & - & 0.03 & 2.32 & 0.29 \\
    \hline
    \end{tabular}
    \label{table:taufitParam}
\end{table*}

\section{Models and simulation details} \label{sec:simuDet}
NVT molecular dynamics simulations using the Brown and Clarke thermostat \cite{Brown1984} were done for the following two dimensional glass-former models. The simulation details are the same as in Ref. \cite{Sengupta2012}.

\subsection{2D MKA model}
The binary mixture ($A_{80} B_{20}$) of Lennard Jones particles introduced by Kob and Andersen (KA) \cite{Kob1995} is a model glass forming liquid in three dimensions. In the case of the modified KA (MKA) model in two dimensions (2D), the composition is set to $A_{65} B_{35}$ \cite{Bruning2008}. The interaction potential, including correction terms that make both potential energy and force to go to zero smoothly at the cutoff, is given by: 
\begin{equation}
V_{\alpha \beta}(r)=
\begin{cases}
         4\epsilon_{\alpha \beta} \left[(\frac{\sigma_{\alpha \beta}}{r})^{12} - (\frac{\sigma_{\alpha \beta}}{r})^{6} \right] + 4 \epsilon_{\alpha \beta} \left[c_0 + c_2 (\frac{r}{\sigma_{\alpha \beta}})^2 \right], \\
         \hspace{8mm} r < \mbox{cutoff} \\
        0 ,  \hspace{0.5cm} r > \mbox{cutoff}
\end{cases}
\end{equation}
where $\alpha, \beta \in \{A, B\}$. Units of length, energy are $\sigma_{AA}$, $\epsilon_{AA}$ respectively. $\epsilon_{AB} = 1.5$, $\epsilon_{BB} = 0.5$, $\sigma_{AB} = 0.80$, $\sigma_{BB} = 0.88$.  The cutoff is at $2.5 \, \sigma_{\alpha \beta}$. Simulations were done at a fixed number density $\rho = 1.20 \, \sigma_{AA}^{-3}$ with system size $N = 2000$ for a temperature range ${k_B T \over \epsilon_{AA}}= 0.36 - 2.00$. The MD integration time step was in the range $dt / (\sqrt{\sigma_{AA}^2 / \epsilon_{AA}}) = 0.003 - 0.005$, with smaller $dt$ values used at higher temperatures. For this model the onset temperature is $T_{onset} = 0.80$. It serves as the reference temperature to differentiate between the low and the high $T$ regimes. For $T >0.90$ (reduced units), only single trajectory has been chosen and for $T < 0.90$, 4-5 independent trajectories were analyzed. The runlength for all trajectories analyzed is more than $100 \, \tau_{\alpha}$.
    
\subsection{2D R10 model}
It is a 50:50 binary mixture of purely repulsive soft disks in two dimensions \cite{Karmakar2010}, with a inverse power law exponent 10. The interaction potential is given by,
\begin{equation}
V_{\alpha \beta}(r) = 
\begin{cases}
 \epsilon \left[(\frac{\sigma_{\alpha \beta}}{r})^{10} \right] + \epsilon \left[ c_{0} +  c_2 (\frac{r}{\sigma_{\alpha \beta}})^{2} + c_4 (\frac{r}{\sigma_{\alpha \beta}})^{4} \right], \\
 \hspace{5mm} r < \mbox{cutoff} \\
     0 , \;\; r > \mbox{cutoff}
\end{cases}
\end{equation}
where $\alpha, \beta \in \{A, B\}$. $\epsilon = 1.0$, $\sigma_{AA} = 1.0$ and $\sqrt{\frac{\sigma_{AA}^2}{\epsilon_{AA}}}$ set the units of energy, length and time respectively.  $\sigma_{BB}$ = 1.40, $\sigma_{AB}$ = 1.18 $\neq$ $\frac{\sigma_{AA}+\sigma_{BB}}{2}$, $c_0 = -0.806140903539923$, $c_2= 0.7$, $c_4 = -0.156300219287607$ represent the correction to make the potential vanish at cutoff$=1.385418025\,\sigma_{\alpha \beta}$ continuously upto the second derivative. Simulations were done at a number density of $\rho \sigma_{AA}^3 = 0.85$ over a range of temperatures ${k_B T \over \epsilon_{AA}}= 0.46 - 2.00$, for a system size of $N=2048$. The onset temperature is at $T_{onset} = 0.80$ (reduced units) which demarcates the high and low temperature regimes. For ${k_B T \over \epsilon_{AA}} \geq 1.00$, a single trajectory has been chosen with integration time step $dt / (\sqrt{\sigma_{AA}^2 / \epsilon_{AA}}) = 0.003$ and for $T \leq T_{onset}$ 3-5 independent trajectories were used with $dt / (\sqrt{\sigma_{AA}^2 / \epsilon_{AA}}) = 0.005$. The runlength for all trajectories analyzed is more than $100 \, \tau_{\alpha}$. 

\subsection{Simulations for anharmonic correction by method 2}
To compute the anharmonic correction by method 2 described in Sec. \ref{sec:Defn}, MD runs were performed for short time duration $\sim$ MSD plateau time. These runs were started from inherent structures obtained from various parent temperatures $T_p$. The range of $T_p$ values were chosen to be $T_P=0.36-0.90$ for the 2DMKA model and $T_p=0.46-0.80$ for the 2DR10 model respectively. Number of runs were 26-926 for the 2DMKA model and 1000-3985 for the 2DR10 model depending on the temperature. For the same initial configuration, runs were performed at various target temperatures $T$ from $10^{-6}$ upto $T_p$. At all target temperatures $T$, runlengths were chosen to be for 1000 $MD$ steps for the 2DMKA model and $3000$ MD steps for the 2DR10 model. Integration time step was set to $dt / (\sqrt{\sigma_{AA}^2 / \epsilon_{AA}}) = 0.003$ for both models.

\section{Results} \label{sec:results}

\begin{figure*}[htp!]
    \centering
    \includegraphics[keepaspectratio, width=0.36\textwidth]{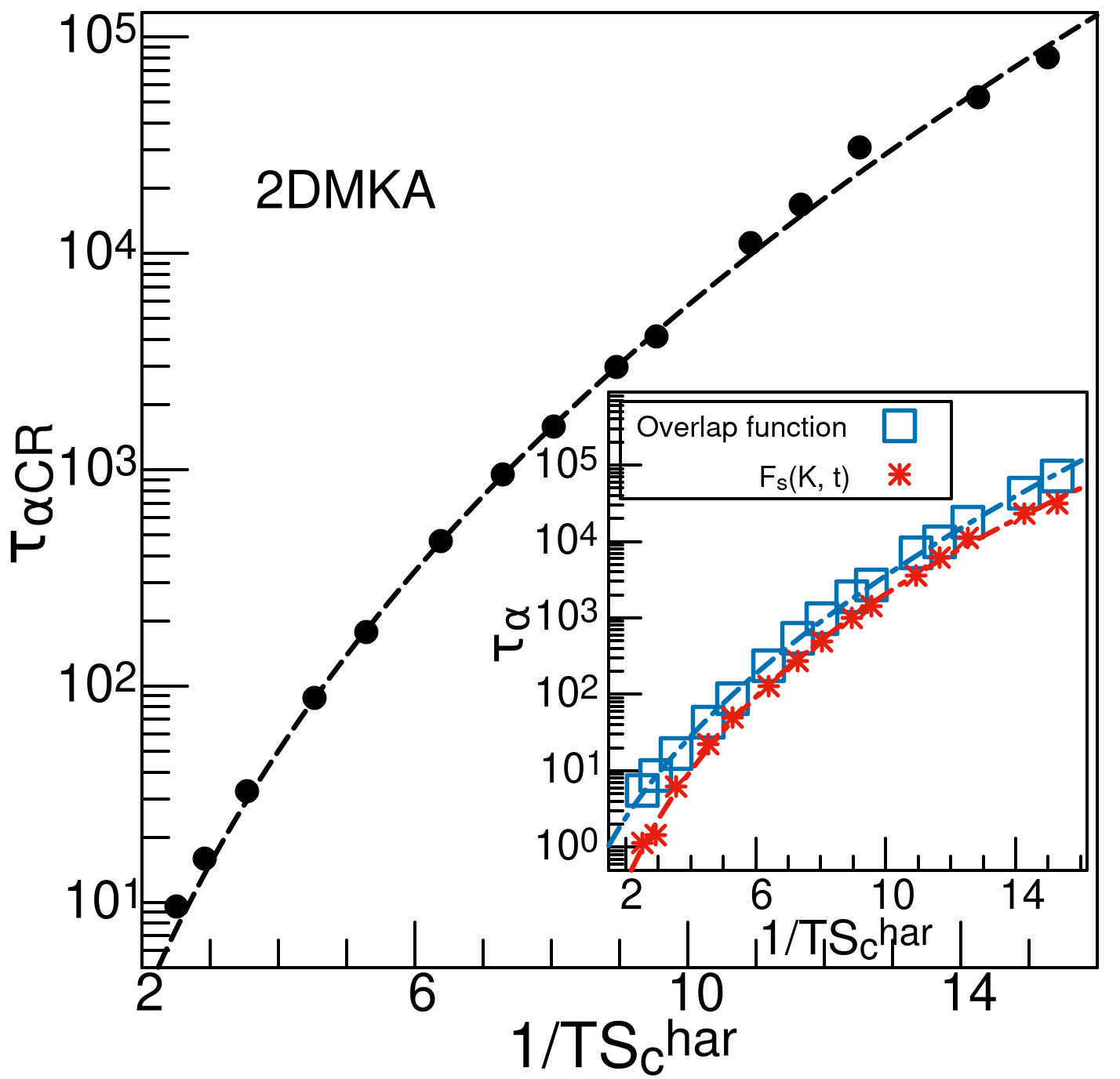}
    \includegraphics[keepaspectratio, width=0.36\textwidth]{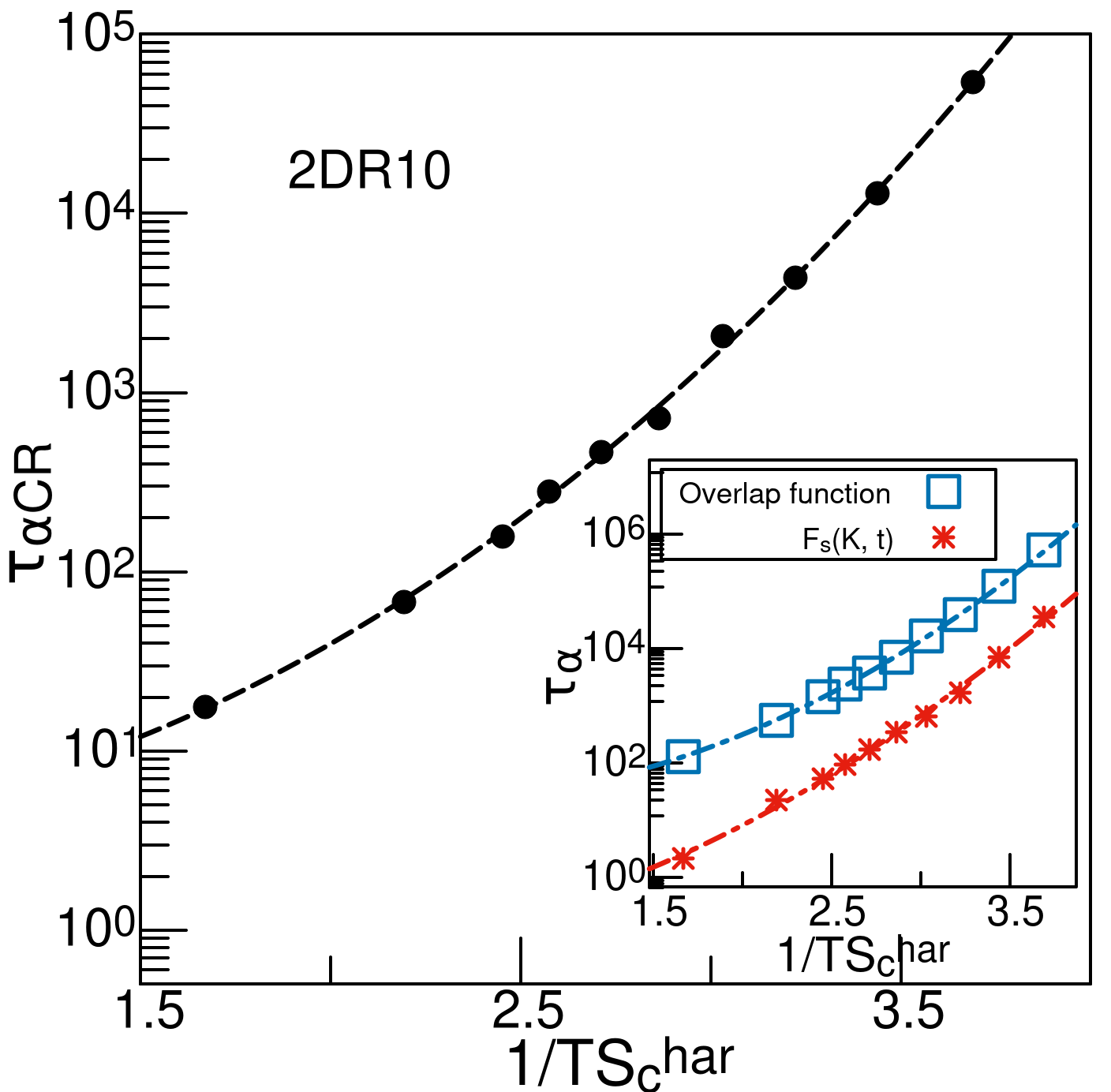}
    \caption{\emph{Effect of long wavelength fluctuations on the Adam-Gibbs (AG) relation.} \emph{(Inset panels):} The AG plots without any correction, {\it i.e.} relaxation time $\tau_{\alpha}$ is computed from correlation functions of standard displacement field and configuration entropy $S_c^{har}$ computed within harmonic approximation. Data for $S_c^{har}$ and $\tau_\alpha$ from overlap function was taken from Ref. \cite{Sengupta2012}. In the present work, we have added the $\tau_\alpha$ data obtained from $F_s(K,t)$.  
    \emph{(Main panels):} The AG plots with relaxation time ($\tau_{\alpha CR}$) corrected for the long wavelength fluctuations {\it vs.} $\frac{1}{TS_c^{har}}$. We show configuration entropy in harmonic approximation to compare with the reference data and isolate the effect of long wavelength fluctuations. Lines are fits to the  generalized Adam-Gibbs's relation $\ln{\tau_\alpha} = \ln \tau_0 + A\left(\frac{1}{TS_c}\right)^{\alpha}$. See text for further details and Table \ref{table:alpha} for fit parameter values.}
    \label{fig:AGCR}
\end{figure*}

\subsection{Comparison of standard {\it vs.} cage-relative dynamics}
In Figs. \ref{fig:Fskt2DMKA} and \ref{fig:Fskt2DR10} we compare the dynamics for the 2DMKA and the 2DR10 models respectively using two estimators: the intermediate scattering functions and the MSD. The estimators computed from standard displacement field are shown in the top rows and those from the cage-relative coordinates are displayed in bottom rows. Both standard and cage-relative estimators show qualitative signatures of glassy dynamics \cite{KobBinderBook}. However, we note the \emph{quantitative} differences: (a) the plateau regime is more prominent for cage-relative $F_s(K,t)$ and MSD, (b) plateau height $f_c$ of the cage-relative intermediate scattering function is higher and (c) the transient wiggle in the plateau regime of the cage-relative $F_s(K,t)$ due to finite size effect is absent in both the models. These observations are in sync with previous studies \cite{Shiba2016}.

Next we study the effect of long wavelength fluctuations on the slowdown of dynamics with reducing temperature. First we compute the usual $\tau_\alpha$ from the dynamics of the standard displacement field in Fig. \ref{fig:compareTime} left columns. Note that Ref. \cite{Sengupta2012} used the overlap function while the present study analyzes the $F_s(K,t)$. We show for both the models that $\tau_\alpha$ extracted from either correlation function are mutually proportional. More interestingly, we extract $\tau_{\alpha CR}$ from the cage-relative intermediate scattering function. We see that the main effect of the long wavelength fluctuation is quantitative: the cage-relative displacement field has longer $\alpha$-relaxation timescale. This finding is consistent with previous observations \cite{Illing2017}. However, the $T$ dependence of the timescales are similar, as the VFT fit $\tau_\alpha (T) = \tau_0 \, \exp({D \over T - T_0})$ to $\tau_\alpha$ from both the $F_s(K,t)$ and the CR-$F_s(K,t)$ yield similar divergence temperature $T_0$ and kinetic fragility ${T_0 \over D}$, see Table \ref{table:taufitParam}. For the sake of completeness, we also show the high T Arrhenius fit to CR-$F_s(K,t)$ timescale to mark the onset temperature of non-Arrhenius dynamics. 

In the low $T$ non-Arrhenius regime, the various characteristic timescales such as the $\alpha$-relaxation time and the diffusion coefficient becomes decoupled - a phenomenon usually labelled as the breakdown of the Stokes-Einstein relation (SEB). Consequently, in the context of testing the AG and the RFOT relations an interesting open question is the choice of timescale to characterize dynamics, which the theories does not explicitly specify. Some of us in a previous study \cite{2017Parmar} showed that for the AG relation, the diffusion coefficient is the more fundamental choice of timescale. Thus we also characterize dynamics by extracting the diffusion coefficient $D_{CR}$ from the cage-relative MSD, see middle panels of Fig. \ref{fig:compareTime}. In the right columns of Fig. \ref{fig:compareTime}, we compare the time scales $\tau_{\alpha CR}$ and the $D_{CR}$ and find a decoupling (SEB) at low temperatures. Thus taken together, these observations indicate that the long wavelength fluctuation affects the dynamics in 2D quantitatively, but qualitative features of the dynamics remains the same. 

\begin{figure}[htp!]
    \centering
    \includegraphics[keepaspectratio, width=0.29\textwidth]{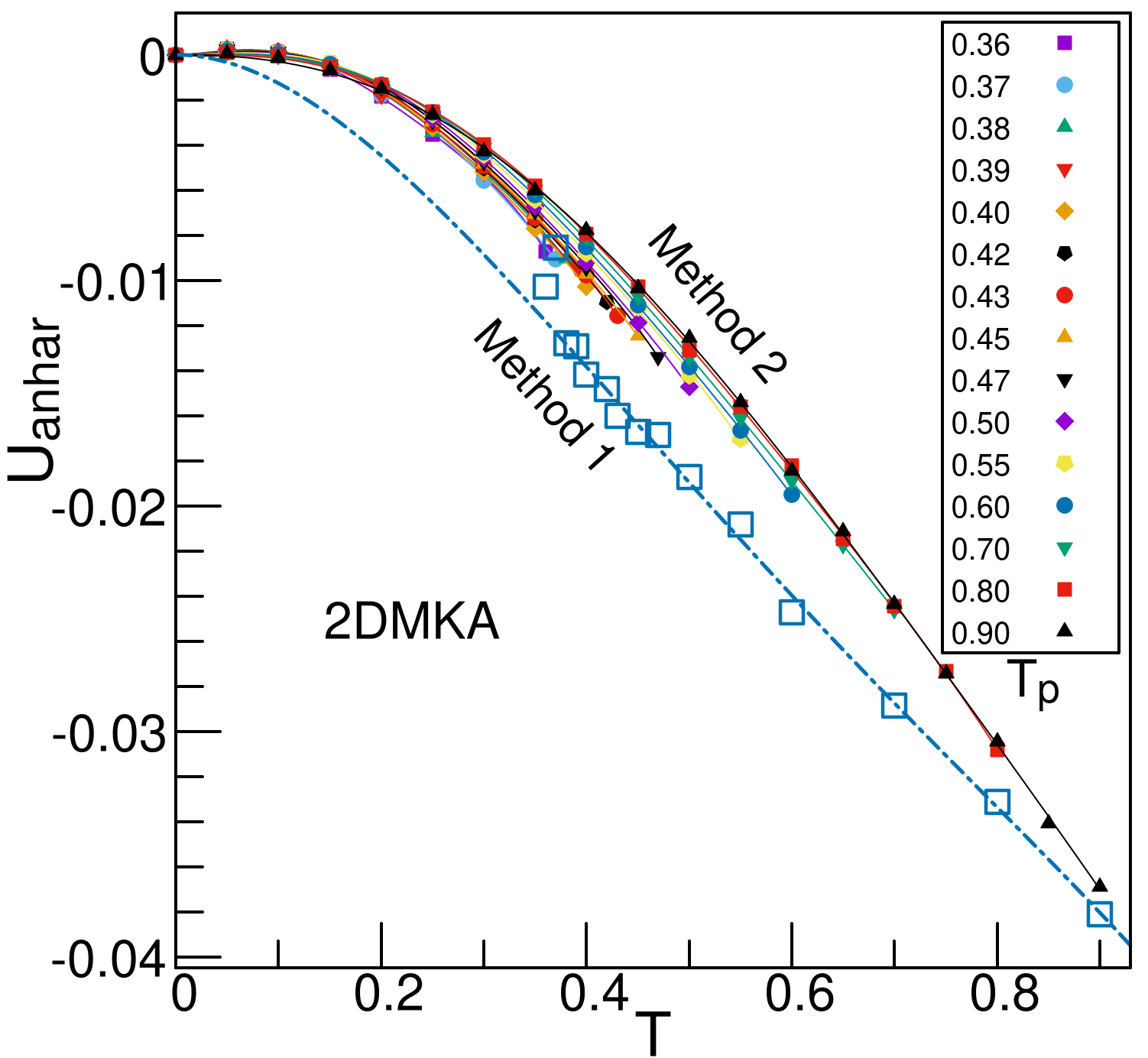}
    \includegraphics[keepaspectratio, width=0.29\textwidth]{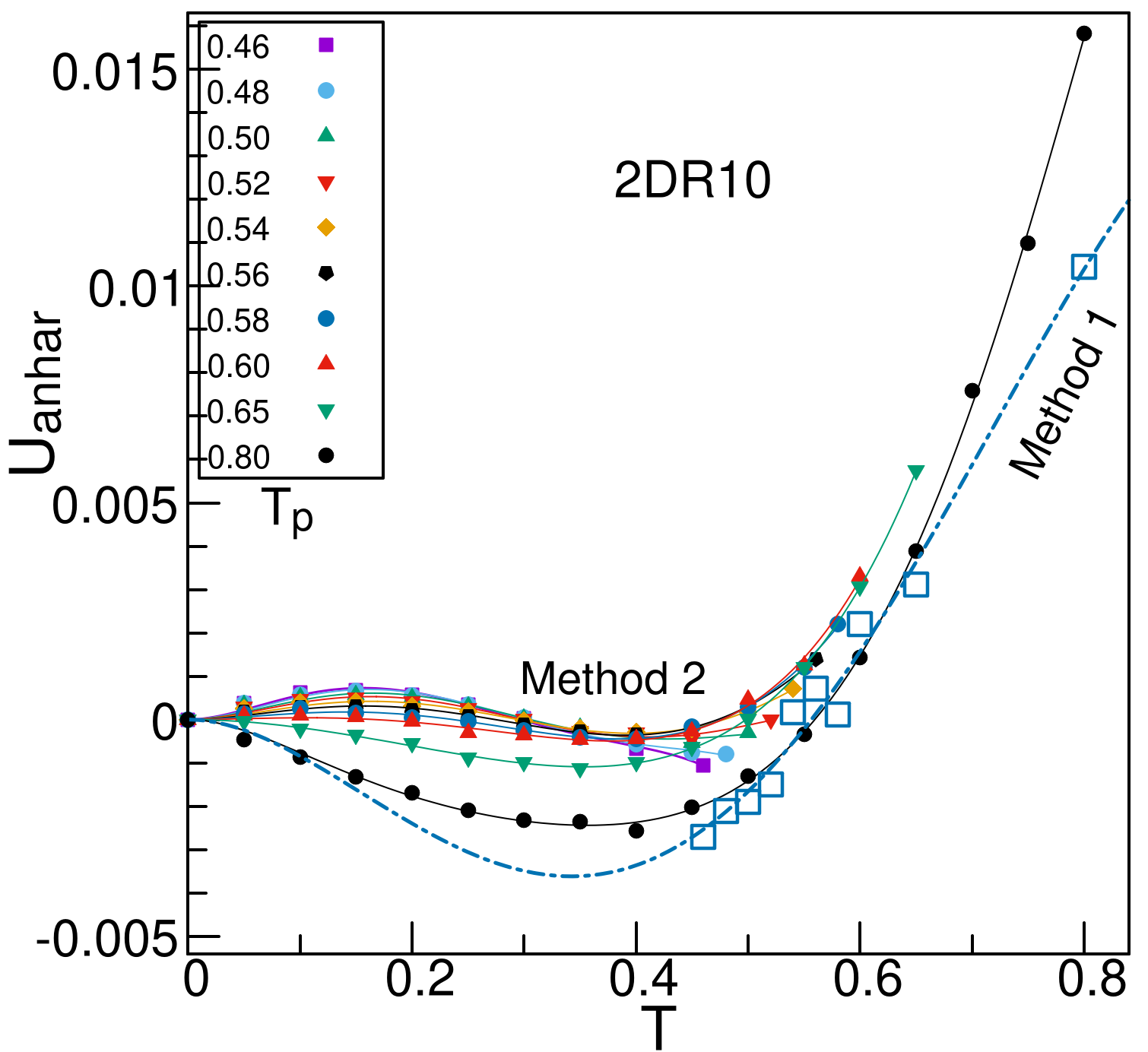}
    \caption{Temperature dependence of anharmonic correction to vibrational potential energy $U_{anhar} (T)$ for the 2DMKA and the 2DR10 models respectively. $U_{anhar} (T)$ is computed using two different methods, see Sec \ref{sec:Defn}. For both models, lines are fits to fourth order polynomials ($j_{max}=4$ in Eqns. \ref{eq:uanh1} and \ref{eq:uanh2}).} 
    \label{fig:uanh}
\end{figure}

\subsection{Effect of the long wavelength fluctuation on the AG relation}
After accounting for the long wavelength fluctuation in dynamics, we pose the next natural question: how does the relation between dynamics and thermodynamics gets affected by such long wavelength fluctuations? To answer it, one is required to test the AG relation without (standard) and with (cage-relative) correction due to the long-wavelength fluctuations. The AG relation predicts a linear dependence of $\ln \tau_\alpha$ on $(TS_c)^{-1}$ at all spatial dimension $D$, Eqn.\ref{eqn:AG}. According to the RFOT theory \cite{Kirkpatrick2015, Lubchenko2007, Bouchaud2004, Biroli2012, Starr2013, Karmakar2015}, the growing many body correlation at lowering temperature is characterized by a growing static correlation length $\xi$. The free energy barrier for structural relaxation $\Delta G$ scales with $\xi$ as $\Delta G \sim \xi^\psi$. Thus the characteristic time scale for structural relaxation scales with $\xi$ as: $\tau_\alpha \sim \exp ({\Delta G \over k_B T}) \sim \exp ({\xi^\psi \over k_B T})$. Further, in the RFOT scenario, the correlation length is related to the configuration entropy as: $\xi^{D - \theta} \sim {1 \over S_c}$. Hence the relation between time and entropy becomes explicitly dimension dependent:
\begin{align}
    \tau_\alpha &\sim \exp \left({A \over (TS_c)^{{\psi \over D-\theta}}}\right)
    \label{eqn:RFOT}
\end{align}
The RFOT exponents $\psi$ and $\theta$ are in general unknown. However, they should satisfy the inequality \cite{Starr2013}: $\theta \leq \psi \leq D-1$. Note that testing the RFOT relations require computation of the lengthscale $\xi$ which is beyond the scope of the present study. However, guided by these scaling relations we fit the data to a non-linear, generalized AG form:
\begin{align}
    \ln{\tau_{\alpha}} = \ln{\tau_0} + A(\frac{1}{TS_c})^{\alpha}
    \label{eqn:genAG}
\end{align}
The AG relation is recovered \cite{Starr2013, Kirkpatrick1989} for $\psi = D/2 = \theta$.

Fig. \ref{fig:AGCR} tests the relationship between the $\alpha$-relaxation timescale and the configurational entropy for the 2DMKA and the 2DR10 models. In all cases the temperature range is chosen upto the onset temperature. Here the configurational entropy is computed using the harmonic approximation, to isolate the effect of long wavelength fluctuation on the AG relation. The inset panels show the timescales $\tau_\alpha$ without correcting for the long wavelength fluctuation, obtained from both the overlap function (Ref \cite{Sengupta2012}) as well as the $F_s(K,t)$.  Comparing the timescales from the intermediate scattering function without (standard) and with (cage-relative) correction due to the long wavelength fluctuation, we see its main effect is \emph{quantitative}: in both cases the AG relation still breaks down, and instead follows the non-linear form in Eqn. \ref{eqn:genAG}. However, removing the effects of long wavelength fluctuation changes the extent of the deviation from linearity, {\it i.e.} values of the exponent $\alpha \equiv {\psi \over D - \theta}$, see Table \ref{table:alpha}. Interestingly, for the 2DMKA model, $\alpha$ increases \emph{towards} 1 wheres for the 2DR10 model, $\alpha$ increases \emph{away} from 1. Thus the nature of deviation depends on the nature of the interatomic potential.  

\begin{figure*}[htp!]
    \centering
    \includegraphics[keepaspectratio, width=0.32\textwidth]{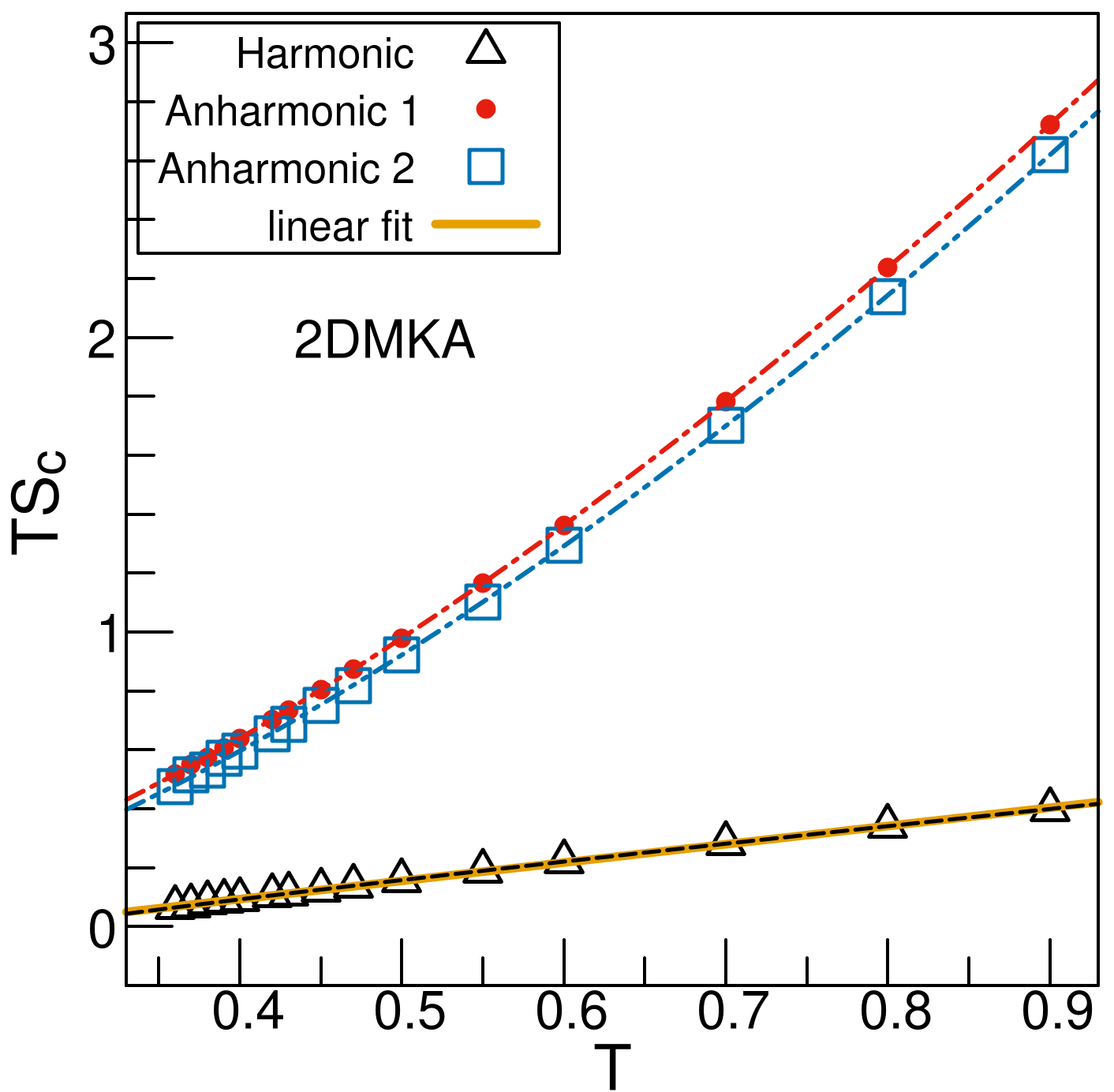}
    \includegraphics[keepaspectratio, width=0.32\textwidth]{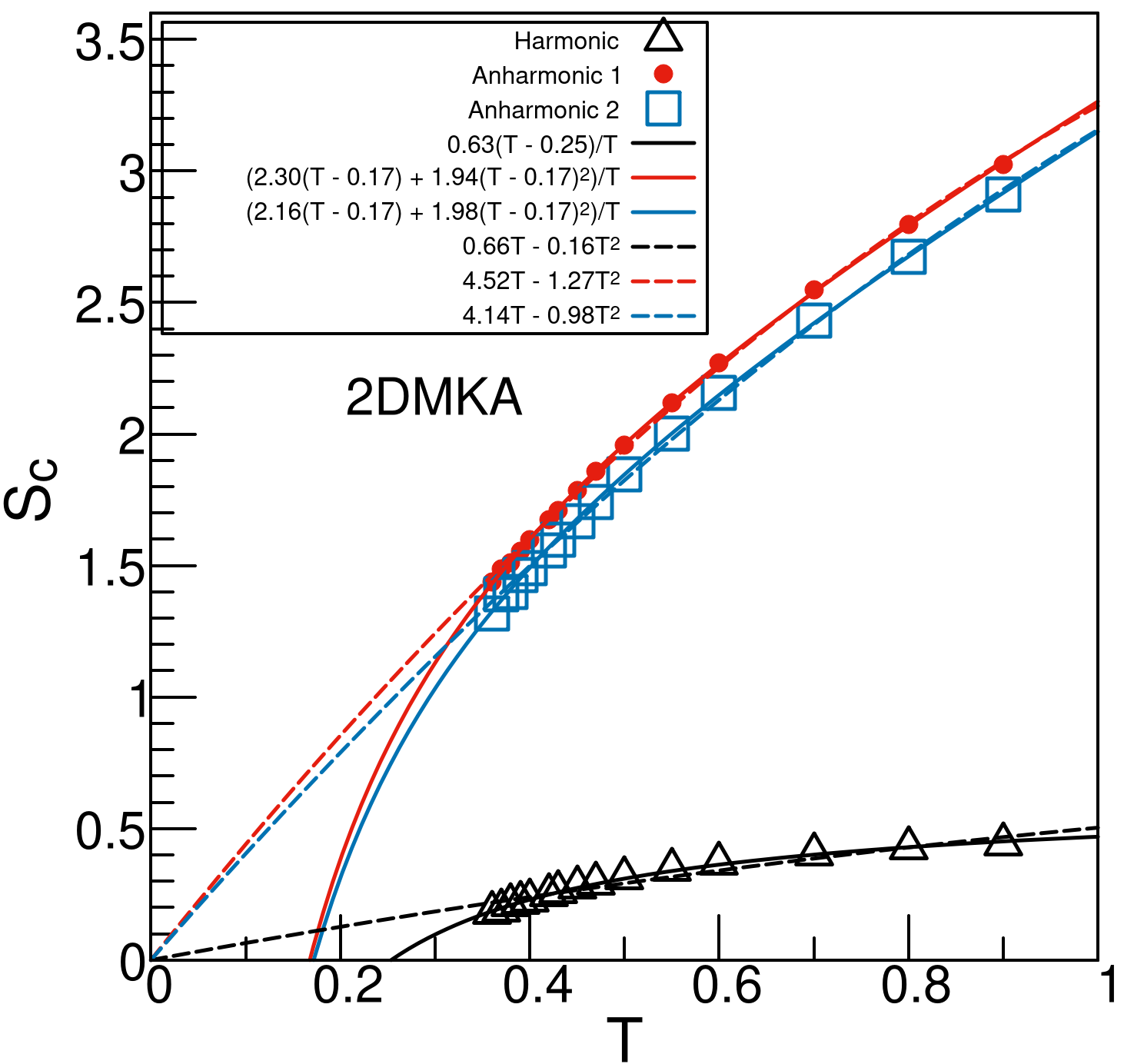}\\
    \vspace{3mm}
    \includegraphics[keepaspectratio, width=0.32\textwidth]{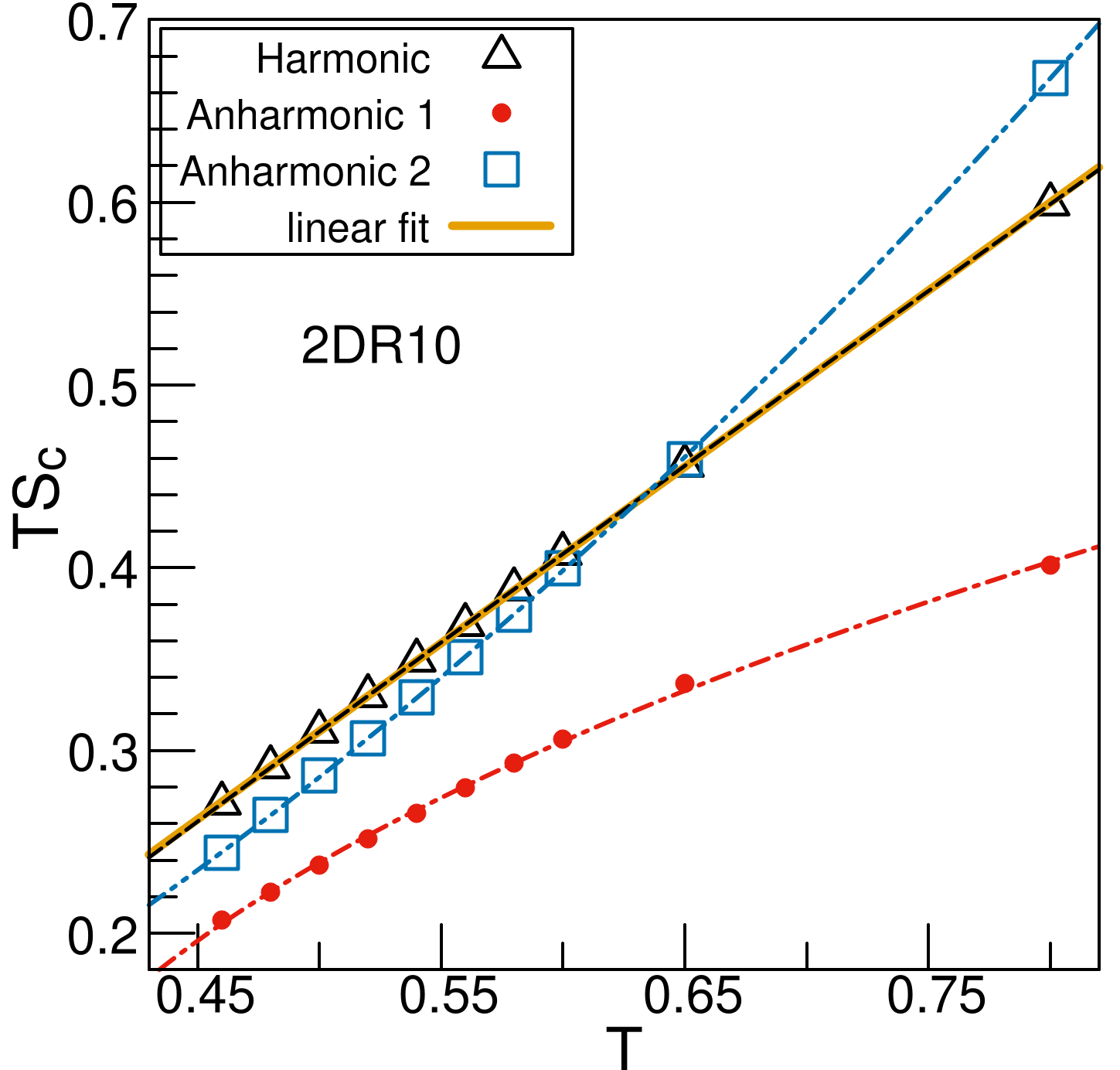}
    \hspace{3mm}
    \includegraphics[keepaspectratio, width=0.32\textwidth]{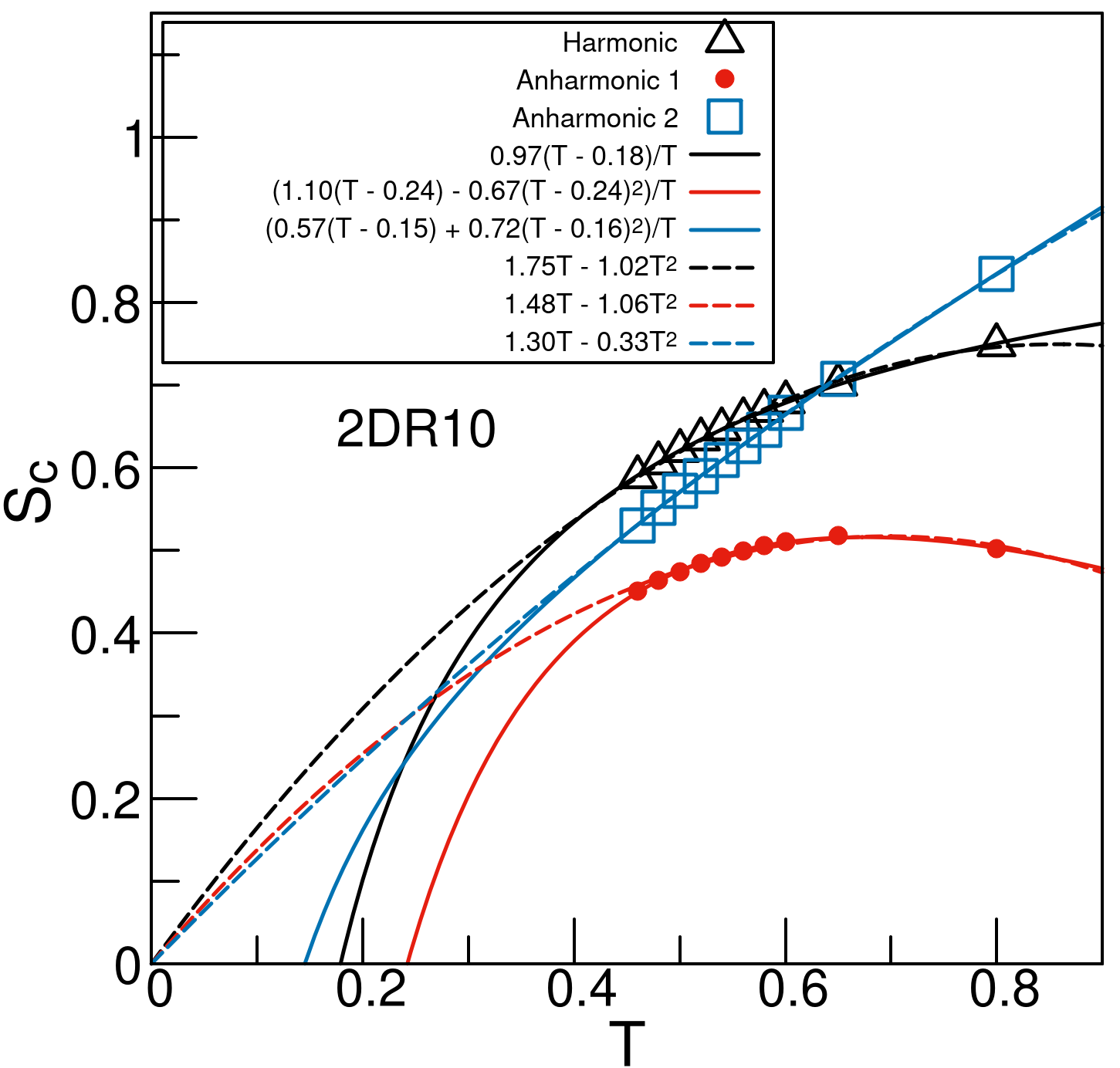}
    \caption{\emph{Effect of anharmonic vibration on the $T$ dependence of configuration entropy.} $TS_c$ vs. Temperature (left column) for the 2DMKA (top row) and the 2DR10 (bottom row) models. $S_c$ is computed in harmonic approximation (black circles, Eqn. \ref{eqn:schar}) as well as by adding anharmonic correction (Eqn. \ref{eqn:scanhar}) using two different methods (Eqns. \ref{eq:uanh1} - \ref{eq:sanh2}). Linear fit ($\alpha=1)$ represent the function: $TS_c = [K_T(\frac{T}{T_K} - 1)]$. To check the deviation from linearity, we fit the data to a generalized non-linear temperature dependence: $TS_c = [K_T(\frac{T}{T_K} - 1)]^{\frac{1}{\alpha_1}}$ where $T_K$ is the Kauzmann temperature, and $K_T$ is the thermodynamic fragility, and $\alpha_1$ is a free parameter. Parameter values are listed in Table \ref{table:TSc}. Right column shows the $T$ dependence of $S_c$ without (black circle) and with anharmonic correction using methods 1 and 2. Fit lines test two types of plausible temperature dependence of $S_c$ - one with $T_K=0$ and the other with $T_K \neq 0$. Our data is consistent with either scenarios. See text for more details.}
    \label{fig:TSc}
\end{figure*}

\begin{table}[htp]
    \centering
    \caption{Parameters for fit to the temperature ($T$) dependence of $TS_c$ for the 2DMKA and the 2DR10 models, Fig. \ref{fig:TSc}. $\alpha_1=1$ represents linear fit $TS_c = [K_T(\frac{T}{T_K}) - 1]$ and $\alpha_1 \neq 1$ sets depict non-linear fit to $TS_c = [K_T(\frac{T}{T_K}) - 1]^{\frac{1}{\alpha_1}}$. Anharmonic 1 and 2 refer to anharmonic correction to $S_c$ computed by methods 1 and 2 respectively.}
    \begin{tabular}{|c|c|c|c|c|}
    \hline
    Model & $S_c$ type & $K_T$ & $T_K$ & $\alpha_1$ \\
    \hline
    \multirow{4}{*}{2DMKA} &  Harmonic  & 0.16 & 0.25  & 1 \\
    \cline{2-5}
         &  Harmonic  & 0.16 & 0.28 & 1.10 \\
    \cline{2-5} 
         & Anharmonic 1 & 0.26 & 0.10 & 0.68 \\
    \cline{2-5}
         & Anharmonic 2 & 0.26 & 0.11 & 0.67 \\
    \hline
    \multirow{4}{*}{2DR10} & Harmonic & 0.17 & 0.18  & 1 \\
    \cline{2-5}
         & Harmonic & 0.19 & 0.20 & 1.05\\
    \cline{2-5}
         & Anharmonic 1 & 0.11 & 0.35 & 2.14\\
    \cline{2-5}
         & Anharmonic 2 & 0.05 & 0.05 & 0.60 \\
    \hline
    \end{tabular}
    \label{table:TSc}
\end{table}

\begin{figure*}[htp!]
    \centering
    \includegraphics[keepaspectratio, width=0.36\textwidth]{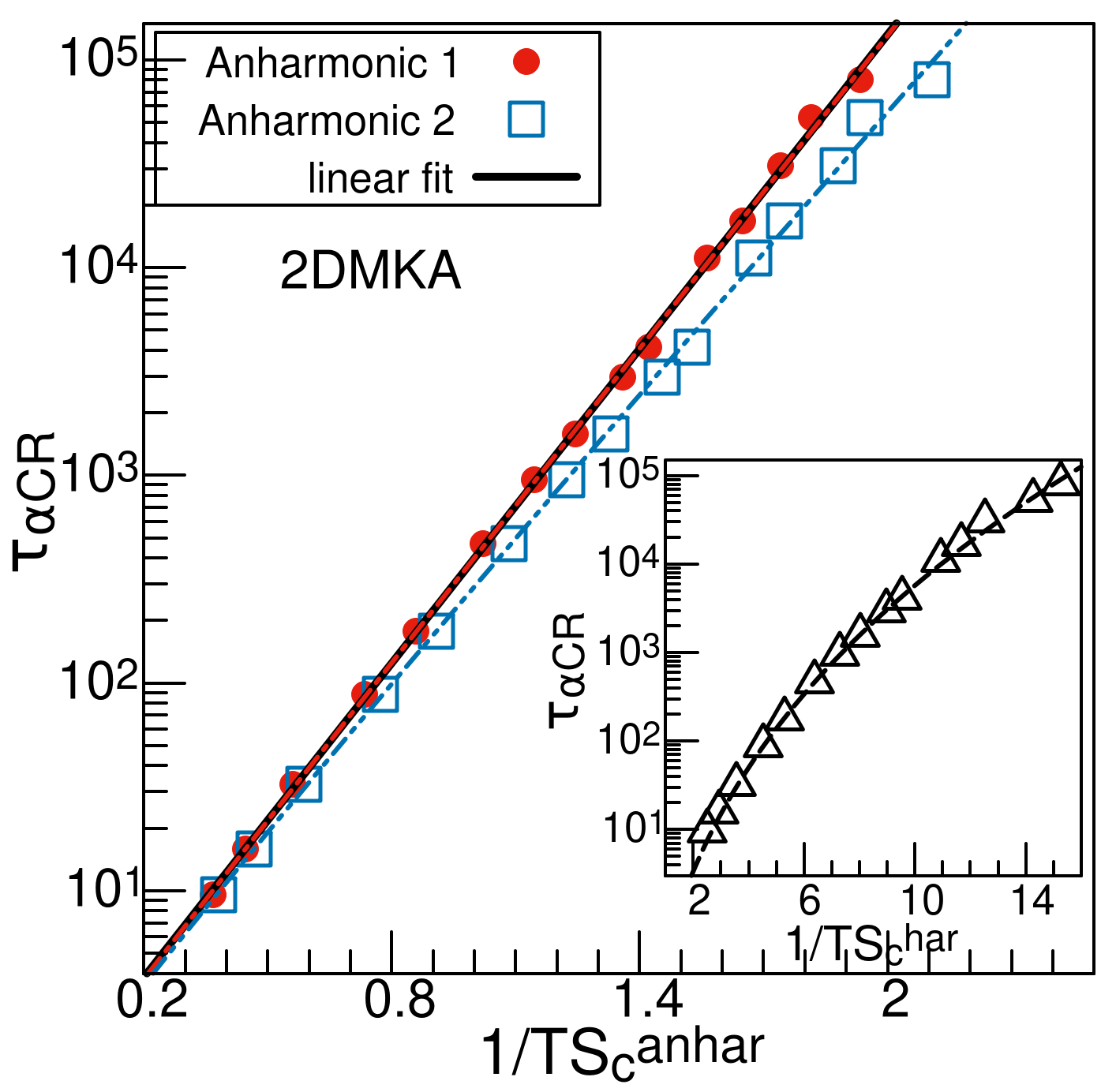}
    \includegraphics[keepaspectratio, width=0.36\textwidth]{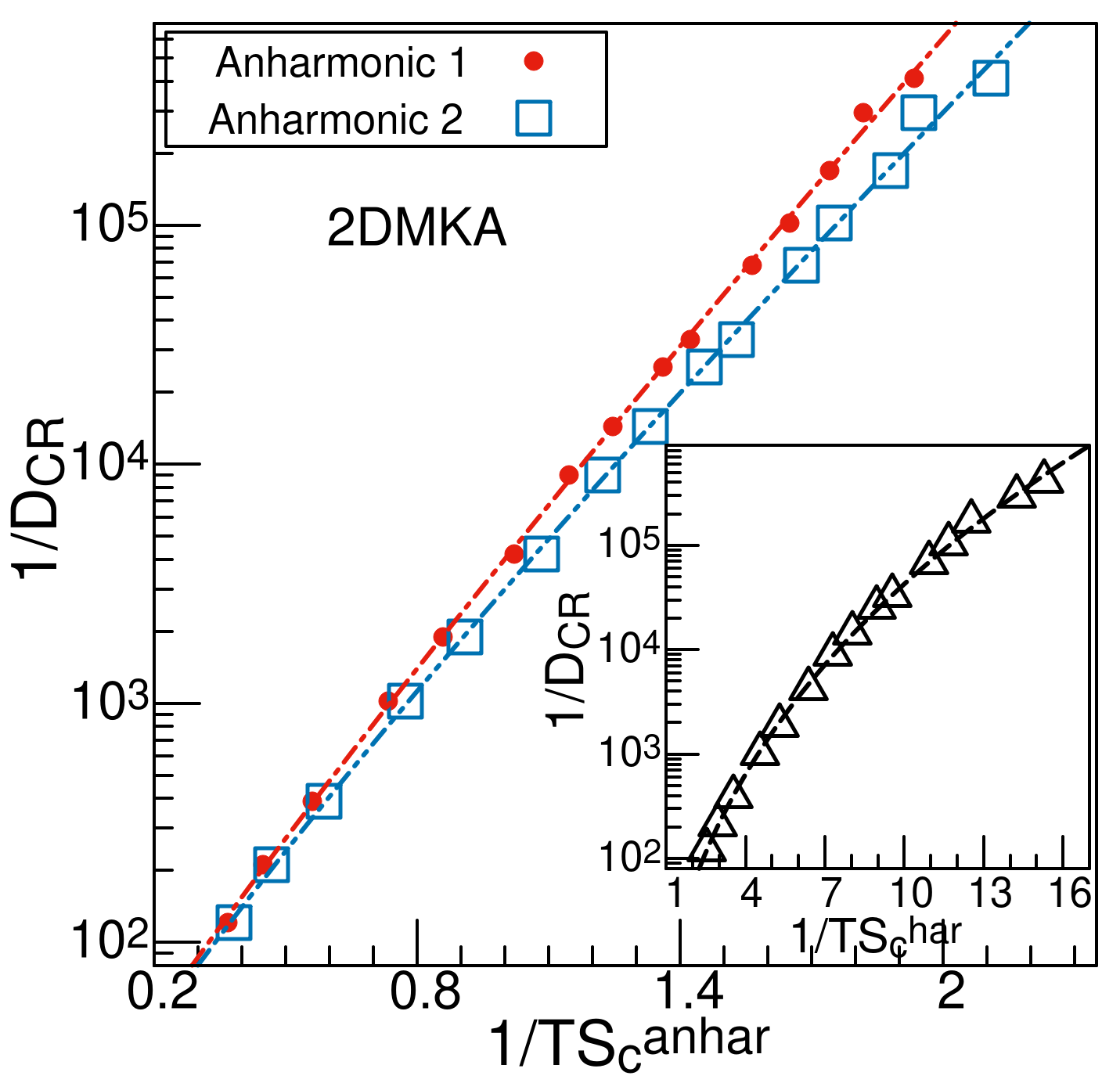}\\
    \includegraphics[keepaspectratio, width=0.36\textwidth]{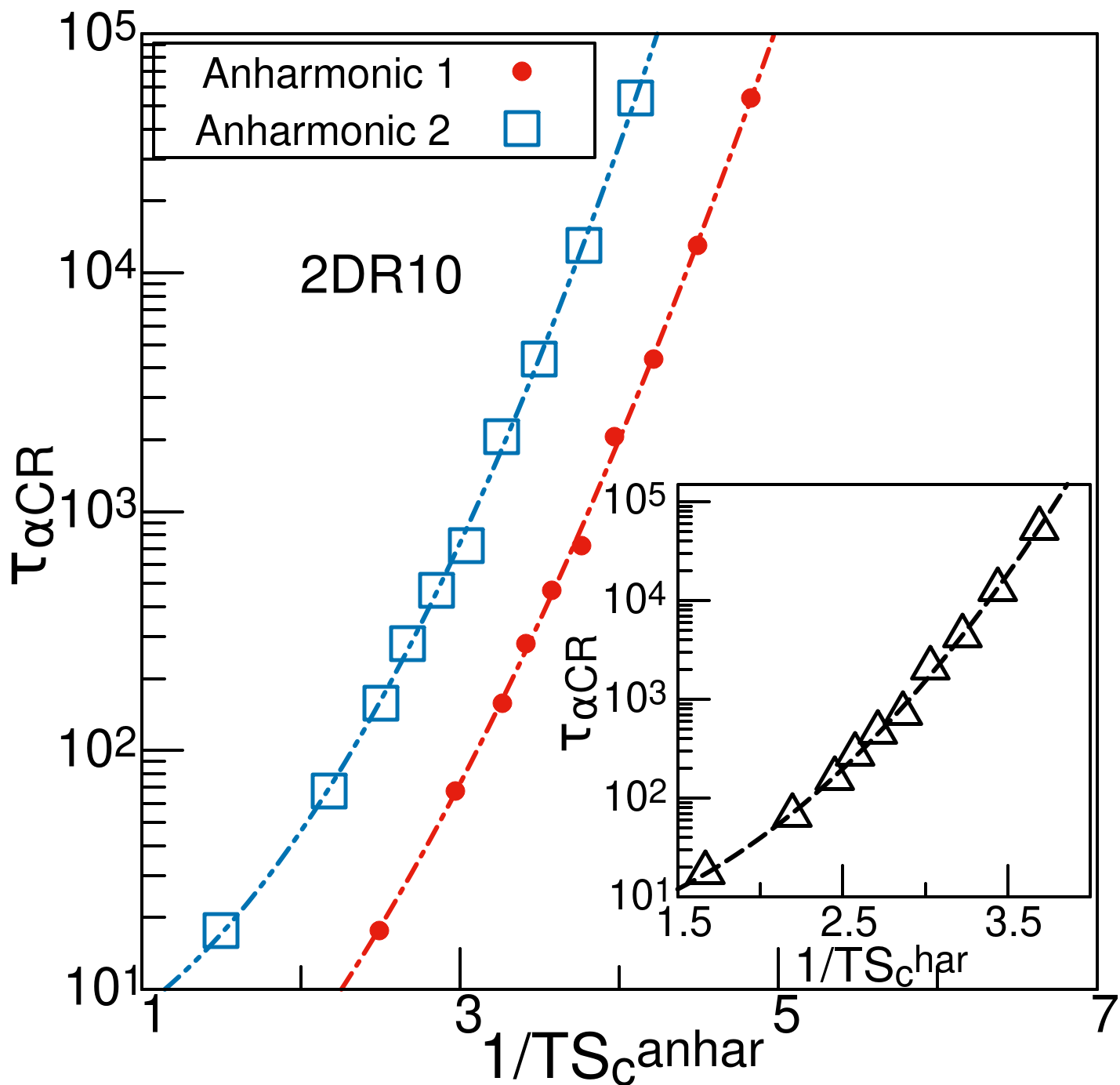}
    \includegraphics[keepaspectratio, width=0.36\textwidth]{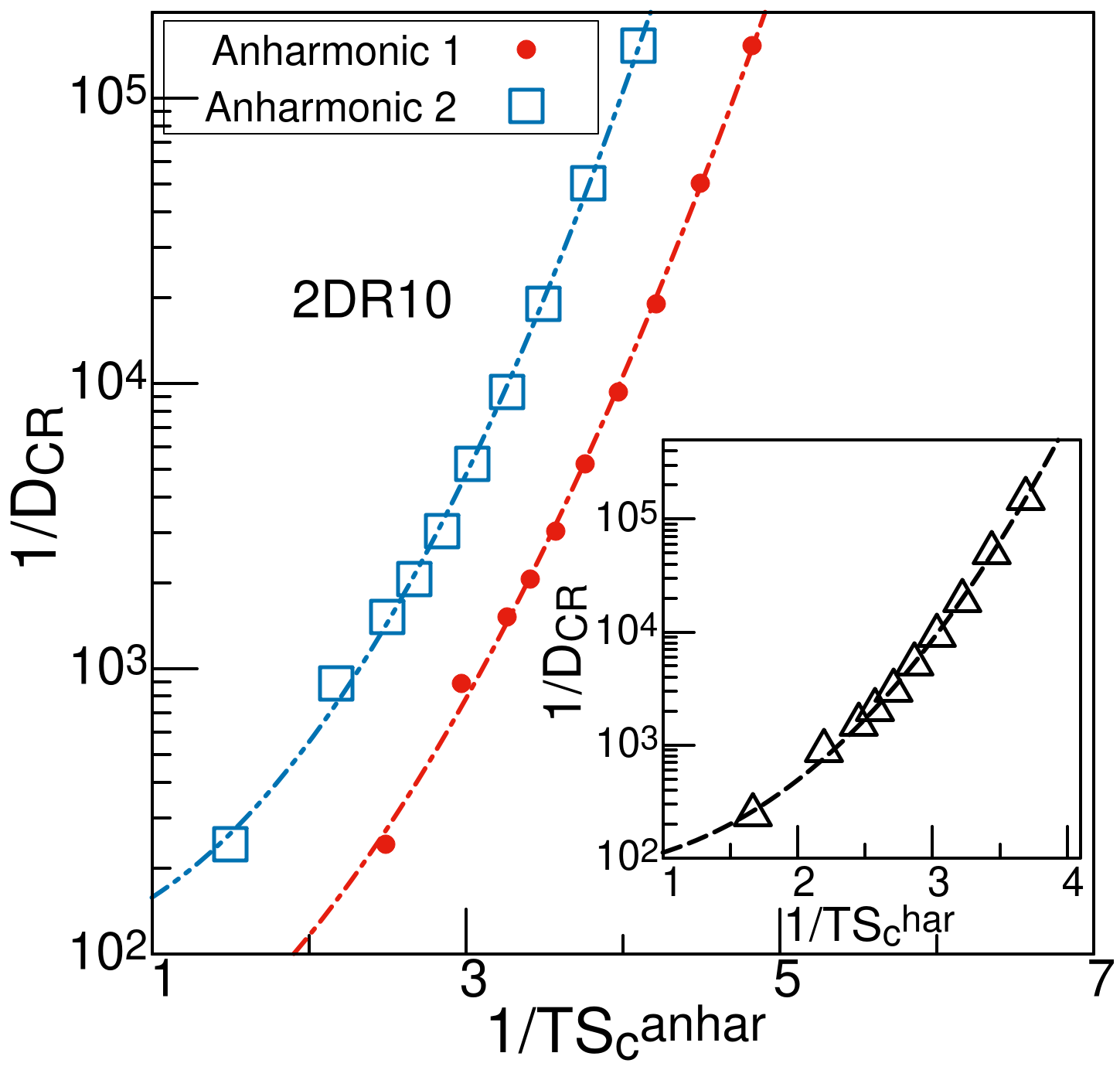}
    \caption{\emph{Effect of anharmonic correction to configurational entropy on the AG relation}: The AG plot in terms of both $\alpha$-relaxation time $\tau_{\alpha CR}$ (left column) and diffusion coefficient $D_{CR}$ for the 2DMKA (top row) and the 2DR10 models (bottom row). The inset shows configuration entropy in harmonic approximation, and the main panel describes $S_c^{anhar}$ where anhamonic vibration effects are explicitly included. $\tau_{\alpha CR}$ is computed from the CR-$F_s(K,t)$. Linear fits represent the AG relation (see Eqn. \ref{eqn:AG}). Non-linear curves are fits to the generalized Adam-Gibbs's relation $\ln \tau_\alpha = \ln \tau_0 + A\left(1/TS_c\right)^{\alpha}$, see Eqn. \ref{eqn:genAG} and Table \ref{table:alpha} for fit parameter values. For the 2DMKA model the AG relation is recovered after incorporating the anharmonic correction, but it still breaks down for the 2DR10 model even after correction due to long wavelength fluctuations and anharmonic vibration are implemented. Thus our data demonstrates intrinsic dimension dependence of the physics of glass transition. See text for further details.}
    \label{fig:AGbothcorr}
\end{figure*}

\begin{table*}[htp]
    \centering
    \caption{Table of the coefficient $A$, exponent $\alpha$ and $\tau_0$ values in the generalized AG relation (Eqn. \ref{eqn:genAG}) for the 2DMKA and the 2DR10 models. 
    }
    \begin{tabular}{ |c|c|c|c|c|c|}
    \hline
    Model & $\tau_{\alpha}$ from & $S_c$ type  & A &$\alpha$ & $\tau_0$ \\
    \hline
    \multirow{7}{*}{2DMKA}     & overlap function & Harmonic  & 4.33 & 0.49 & $5.55 \times 10^{-3}$ \\
    \cline{2-6}
       & $F_s(K,t)$ & Harmonic  & 17.26 & 0.22 & $6.11 \times 10^{-10}$ \\
    \cline{2-6}
       & CR-$F_s(K,t)$ &  Harmonic  & 7.72 & 0.35 & $1.77\times 10^{-4}$ \\
    \cline{2-6}
       & overlap function & Anharmonic 1 & 5.41 & 1.11 & 0.95 \\
    \cline{2-6}
       & $F_s(K,t)$ & Anharmonic 1 & 6.51 & 0.97 & 0.16 \\
    \cline{2-6}
       & CR-$F_s(K,t)$ &  Anharmonic 1 & 5.86 & 0.99 & 1.14 \\
    \cline{2-6}
       & CR-$F_s(K,t)$ &  Anharmonic 2 & 5.64 & 0.95 & 1.03 \\
    \hline
    \multirow{7}{*}{2DR10} & overlap function & Harmonic  & 0.65 & 2.10 & 19.12 \\
    \cline{2-6}
         & $F_s(K,t)$ & Harmonic  & 1.25 & 1.76 & 0.12 \\
    \cline{2-6}
         & CR-$F_s(K,t)$ &  Harmonic  & 0.56 & 2.19 & 3.10 \\
    \cline{2-6}
         & overlap function & Anharmonic 1 & 0.83 & 1.71 & 2.68 \\
    \cline{2-6}
         & $F_s(K,t)$ & Anharmonic 1 & 2.11 & 1.29 & $2.89 \times 10^{-3}$ \\
    \cline{2-6}
         & CR-$F_s(K,t)$ &  Anharmonic 1 & 0.72 & 1.77 & 0.49 \\
    \cline{2-6}
        & CR-$F_s(K,t)$ &  Anharmonic 2 & 0.59 & 1.96 & 4.62 \\
    \hline
    \end{tabular}
    \label{table:alpha}
\end{table*} 

\subsection{Effect of anharmonic vibration on the AG relation}
Details of configurational entropy calculation in harmonic approximation (Eqn. \ref{eqn:schar}) for the 2DMKA and the 2DR10 models have been reported elsewhere \cite{Sengupta2012}. Here we describe the effect of adding anharmonic correction to $S_c$ (Eqn. \ref{eqn:scanhar}). First, in Fig. \ref{fig:uanh} we show the temperature dependence of the anharmonic contribution to the potential energy characterized by polynomial fit functions, see Sec \ref{sec:Defn} and Eqns. \ref{eq:Uanhar}, \ref{eq:uanh1} and \ref{eq:uanh2}. 

 Analysis within the PEL framework has shown that for harmonic approximation to the vibrational entropy, $TS_c$ should be linear in $T$: $TS_c = [K_T(\frac{T}{T_K} - 1)]$, where $S_c(T_K)=0$ determines the Kauzmann temperature $T_K$ and $K_T$ is the thermodynamic fragility \cite{Sastry2001,Novikov2022}. This expectation is well tested for 3D glass-former models in experiments \cite{Richert1998}, and also in simulation studies without \cite{Banerjee2016} and with anharmonic corrections \cite{Nandi2022} although deviations have also been noted \cite{Ozawa2019}. Recently Berthier {\it et al.} has raised the intriguing possibility that the Kauzmann temperature in 2D is, in fact, zero \cite{Berthier2019}. In other words, there is no ideal glass transition at non-zero temperature in 2D. Given these background, the effect of the correction to $S_c$ due to vibration entropy - see Eqns. \ref{eq:sanh1}, \ref{eq:sanh2} - on its $T$ dependence becomes especially interesting in 2D to which we turn our attention next.  

In Fig. \ref{fig:TSc} we show the $S_c$ data over a range of low temperatures upto the onset temperature for 2DMKA (top row) and 2DR10 (bottom row) models. First note that when the configurational entropy is computed in harmonic approximation, left columns of Fig. \ref{fig:TSc}, $TS_c^{har}(T)$ is indeed linear with $T$. However the Kauzmann temperature $T_K$ is siginificantly different from the VFT divergence temperature $T_0$ in both the models, see Tables \ref{table:taufitParam}, \ref{table:TSc}. We also fit the data using a non-linear function $TS_c = [K_T(\frac{T}{T_K} - 1)]^{\frac{1}{\alpha_1}}$ with $\alpha_1$ as a free fit parameter. The choice of this fit function is motivated by Eqn. \ref{eqn:genAG}, although other fit functions have also been used in the literature \cite{Ozawa2019}. 
This analysis results in the exponent $\alpha_1 \sim 1$ within numerical accuracy providing a robust test of the linearity of $TS_c^{har}(T)$.

To the contrary, we find that in 2D glass-formers, adding anharmonic correction makes the $TS_c^{anhar}$ distinctly \emph{non-linear} with $T$. This behaviour is the same in both models and using both methods 1 and 2 for computing the anharmonic vibrational entropy (see Eqns. \ref{eq:uanh1}-\ref{eq:sanh2}). However, estimates of $T_K$ depends on the models and the methods of computing the anharmonic correction to $S_c$. On one hand, in the 2DMKA model, the non-linear fit estimate of $T_K$ values are significantly lower when anharmonic contribution to vibrational entropy is considered, and approximately same for both methods 1 and 2. On the other hand, in the 2DR10 model, method 1 results in $T_K \sim T_0$ while method 2 produces a $T_K$ closer to zero, see Table \ref{table:TSc}. In order to gain further insight, on the right columns of Fig. \ref{fig:TSc}, we show the temperature dependence of $S_c$. We analyze the $T$ dependence using several plausible fitting functions which can be organized into two families: one assuming $T_K = 0$ \cite{Berthier2019} and the other assuming $T_K \neq 0$. However, both families produce fits of comparable quality. Hence in the range of temperature in which configurational entropy is analyzed in the present study, our data is consistent with both finite and zero temperature ideal glass transition scenarios.

Finally, in Fig. \ref{fig:AGbothcorr}, we test the AG relation with (main panel) and without (inset) including the anharmonic correction to the configurational entropy. Since the timescales at low temperatures gets decoupled (see Fig. \ref{fig:compareTime}), we consider both $\alpha$-relaxation time $\tau_{\alpha CR}$ as well as diffusion coefficient $D_{CR}$. Note that the correction due to the effect of long wavelength fluctuations are already incorporated. Thus \emph{Fig. \ref{fig:AGbothcorr} constitutes the main result of the present work}. The top row of Fig. \ref{fig:AGbothcorr} shows data for 2DMKA model and the bottom row depicts 2DR10 model data. Lines are fits to Eqn. \ref{eqn:genAG} from which the scaling exponent $\alpha$ is extracted and tabulated in Table \ref{table:alpha}. First, for the 2DMKA model, comparing the inset and the main panel, we see that the main effect of anharmonic correction is to reduce the deviation from the AG relation, Eqn. \ref{eqn:AG}, {\it i.e.} to increase the value of $\alpha$ towards 1. This behaviour remains same for both $\tau_{\alpha CR}$ and $D_{CR}$. Indeed, for the 2DMKA model, the AG relation is recovered within numerical accuracy when both the correction to $\tau_{\alpha CR}$ due to long wavelength fluctuation and the correction to $S_c$ due to anharmonic vibration are considered. Even with $D_{CR}$, the AG relation is only weakly non-linear for this model. To the contrary, the 2DR10 model shows a qualitatively different behaviour. Even after accounting for all the corrections in timescales and entropy, one still observes a breakdown of the AG relation in two dimensions. Thus we clearly reveal an intrinsic dimension dependence that indicates that the behavior of 2D glass-formers are different from those in 3D. Interestingly, the nature of inter-particle interaction has a more fundamental effect in determining the value of the exponent $\alpha$ than long wavelength fluctuations and anharmonic correction. Although, even in 2DR10 model, including the anharmonic correction reduces the extent of deviation from linearity, {\it i.e.} decreases $\alpha$ towards 1. 

\section{Summary and conclusion} \label{sec:conclude}
In the present work we carefully examine whether the behaviour of glass-formers in 2D is intrinsically different from those in 3D. To this aim we test the validity of the well-known Adam-Gibbs (AG) relation between dynamics and thermodynamics in two different 2D glass-forming models. In particular, we critically examine the role of two factors that affect the dynamics and the thermodynamics respectively. First, we consider the dynamics of cage-relative, local displacement fields to eliminate the effect of Mermin-Wagner type long wavelength fluctuations that is unique to 2D. Second, we explicitly consider the anharmonic nature of vibration and implement anharmonic corrections to the configurational entropy, computed using two different methods. Recent works in literature have emphasized that even for amorphous matter in 2D, it is essential to explicitly take into account these factors before testing any theory of glass transition. 

We demonstrate that even after accounting for these effects, the AG relation still breaks down in 2D, although the details crucially depends on the nature of the inter-particle interactions. This suggests that Mermin-Wagner type fluctuations does not capture all the differences in dynamics between 2D and 3D. Our analysis implies that behavior of glass-formers in 2D are intrinsically different from those in 3D. Since the two effects are independent of each other, we are able to compare the relative importance of the two factors for the validity of the AG relation. We find that the role of long wavelength fluctuations is mainly quantitative - it does not change the scaling relation between timescale and entropy qualitatively. To the contrary, the effect of anharmonicity of vibration is more significant in the temperature range we study because it can affect the AG relation in both qualitative and quantitative ways. We also verify that our data is consistent with both zero and non-zero temperature glass transition scenarios. This probably indicates that our analysis spans a temperature range that is not sufficient to resolve this issue, and one requires to sample lower temperature in equilibrium. 

The question naturally arises about the origin of the observed deviation from the AG relation in 2D. Since we consider a system (2DMKA) with Lennard-Jones type interaction having an attractive component as well as a glass-former (2DR10) with purely repulsive interaction, we can systematically check a third factor - the role of inter-particle interactions. We show that the nature of deviation depends on the attractive {\it vs.} repulsive nature of the interaction potential. Whether a more advanced theory such as the Random First Order Transition (RFOT) theory can explain the breakdown of the AG relation relation in 2D is an intriguing open question. We note that the observed non-linearity of the scaling relation between timescale and entropy is consistent with RFOT theory, and our analysis implies that the RFOT scaling exponents should be material specific and not universal. However, to completely clarify this issue, one requires to compute the RFOT exponents, which we leave for a future study. 

\bibliographystyle{unsrt}
\bibliography{bibliog.bib}

\end{document}